\documentclass[journal]{new-aiaa}
\usepackage[utf8]{inputenc}
\usepackage{textcomp}
\usepackage{xcolor}
\usepackage{graphicx}
\usepackage{amsmath}
\usepackage{comment}
\usepackage[version=4]{mhchem}
\usepackage{siunitx}
\usepackage{longtable,tabularx}
\usepackage{multirow}
\usepackage{arydshln}
\usepackage{mathtools}
\usepackage{setspace}
\usepackage[font={normalsize,bf}]{subcaption}

\DeclarePairedDelimiter\norm{\lVert}{\rVert}%
\makeatletter
\let\oldnorm\norm
\def\norm{\@ifstar{\oldnorm}{\oldnorm*}}
\makeatother

\title{Seamless Active Morphing Wing Simultaneous Gust and Maneuver Load Alleviation}

\author{Xuerui Wang\footnote{Assistant Professor, Department of Aerospace Structures and Materials, and Department of Control and Operations, Faculty of Aerospace Engineering, Kluyverweg 1, 2629HS Delft, the Netherlands, X.Wang-6@tudelft.nl, AIAA Member.}, Tigran Mkhoyan\footnote{Ph.D. Candidate, Department of Aerospace Structures and Materials, Faculty of Aerospace Engineering, Kluyverweg 1, 2629HS Delft, the Netherlands, T.Mkhoyan@tudelft.nl, AIAA Student Member.}, Iren Mkhoyan\footnote{Researcher, Department of Aerospace Structures and Materials, Faculty of Aerospace Engineering, Kluyverweg 1, 2629HS Delft, the Netherlands, A.Mkhoyan@tudelft.nl.} and Roeland De Breuker\footnote{Associate Professor, Department of Aerospace Structures and Materials, Faculty of Aerospace Engineering, Kluyverweg 1, 2629HS Delft, the Netherlands, R.DeBreuker@tudelft.nl, AIAA Senior Member.}}
\affil{Delft University of Technology, Faculty of Aerospace Engineering, \\ Kluyverweg 1, 2629 HS Delft, The Netherlands}

\begin{document}
\newtheorem{theorem}{Theorem}
\newtheorem{lemma}{Lemma}
\newtheorem{corollary}{Corollary}
\newtheorem{remark}{Remark}
\newtheorem{definition}{Definition}
\newtheorem{assumption}{Assumption}
\newtheorem{proposition}{Proposition}

\maketitle

\begin{abstract}

This paper deals with the simultaneous gust and maneuver load alleviation problem of a seamless active morphing wing. The incremental nonlinear dynamic inversion with quadratic programming control allocation and virtual shape functions (denoted as INDI-QP-V) is proposed to fulfill this goal. The designed control allocator provides an optimal solution while satisfying actuator position constraints, rate constraints, and relative position constraints. Virtual shape functions ensure the smoothness of the morphing wing at every moment. In the presence of model uncertainties, external disturbances, and control allocation errors, the closed-loop stability is guaranteed in the Lyapunov sense. Wind tunnel tests demonstrate that INDI-QP-V can make the seamless wing morph actively to resist ``1-cos'' gusts and modify the spanwise lift distribution at the same time. The wing root shear force and bending moment have been alleviated by more than 44~\% despite unexpected actuator fault and nonlinear backlash. Moreover, during the experiment, all the input constraints were satisfied, the wing shape was smooth all the time, and the control law was executed in real time. Furthermore, as compared to the linear quadratic Gaussian (LQG) control, the hardware implementation of INDI-QP-V is easier; the robust performance of INDI-QP-V is also superior. 
\end{abstract}

\section{Introduction}

\lettrine{T}{he} advancements in aerospace engineering, paired with continuing desire to develop more fuel-efficient aircraft, lead to increasingly flexible aircraft designs. Generally, the flexibility is considered as a side effect of the lighter aircraft design and needs to be adequately accounted for to prevent undesired aerodynamics-structure couplings and ensure the optimized aerodynamic shape. While the flexibility can be accounted for with either passively tailored structural design or active control mechanisms, a fixed-wing shape - generally optimized for the cruise condition - cannot be fully optimized throughout the flight envelope due to conflicting requirements~\cite{Weisshaar2013a}. A more natural approach is to \emph{utilize} the flexibility and actively change the shape by in-flight morphing. This allows the wing to continuously adapt to the most optimal shape when transitioning from one flight phase to the other. Secondly, as compared to the conventional discrete trailing-edge surfaces, smooth morphing can execute flight control and load alleviation commands with reduced noise and drag. The combination of these two aspects can contribute to a more efficient flight routine and a reduced structural weight, thereby increasing the flight sustainability.

\begin{figure}[!h]
\centering
\includegraphics[width=.6\textwidth]{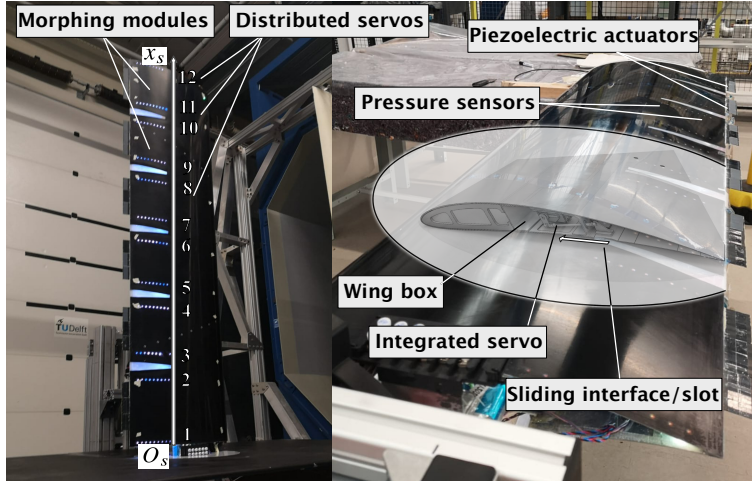}
\caption{The seamless active morphing wing SmartX-Alpha with the TRIC morphing mechanism.}
\label{fig_morphing_mechanism}
\end{figure}

Many research efforts have been devoted to morphing, including piezoelectricity, shape memory alloys materials, compliant actuation mechanism, etc.~\cite{Hubbard2006}. Among all these research aspects, actuation force reduction is one of the bottlenecks of morphing realizations. In~\cite{Previtali2014a}, the actuation force is reduced by a compliant skin mechanism, and a combination between conventional and piezoelectric actuation. However, this approach results in significant manufacturing challenges and complexities. Other concepts, such as the fish bone active camber (FishBAC)~\cite{Woods2015} and the mission adaptive digital composite aerostructure technologies (MADCAT)~\cite{Cramer2019a} demonstrate morphing with ultralight structures. However, the majority of the wing volume is consumed for morphing mechanisms, leaving limited room for other components. Overviewing the state of the art, the key shortcomings of existing morphing techniques include 1) restricted morphing motions; 2) manufacturability and scalability complexities; 3) compromised internal wing volume; 4) inadmissibility for distributed morphing control along the wing span. To overcome these shortcomings, a distributed seamless active morphing wing concept is proposed in~\cite{Tigran_smasis}. As shown in Fig.~\ref{fig_morphing_mechanism}, this morphing wing named SmartX-Alpha is based on the translation induced camber (TRIC) concept~\cite{Tigran_smasis}, which means a cut is introduced to allow the bottom skin to slide in cord-wise and transverse directions. By altering the actuation directions, a pair of actuators can introduce pure camber morphing or warp-induced spanwise twist morphing. To ensure the seamlessness, the adjacent TRIC modules are connected with elastomeric skin, whose stiffness is designed considering the aerodynamic shape holding and the actuation loads. The control algorithms proposed in this paper will be applied to the SmartX-Alpha morphing wing\footnote{The project video can be found via \url{ https://www.youtube.com/watch?v=SdagIiYRWyA&t=319s}}. 

Maneuver load alleviation (MLA) and gust load alleviation (GLA) are two important objectives in aircraft control. Conventional MLA relies on some pre-designed control logic, e.g., when the measured load exceeds a pre-defined threshold, the pre-selected wing control surfaces are triggered to deflect and hold for a certain time period~\cite{fabre1993system}. However, this conventional approach is not efficient and also requires intensive tuning efforts. By contrast, Ref.~\cite{Pereira2019} uses linear model predictive control (MPC) and linear quadratic regulator (LQR) to satisfy the load constraints at various critical stations. In~\cite{Haghighat2012}, the nonlinear flexible aircraft model is linearized successively, and then the MPC controllers are designed at every linearization point. A linear quadratic Gaussian (LQG) control is designed for a SensorCraft vehicle GLA problem in~\cite{Vartio2008a}. Besides, a wind tunnel experiment for alleviating the gust loads of a flexible wing with piezoelectric control is presented in~\cite{Ying2016}. The piezoelectric patches are actuated by a proportional-integral-derivative (PID) controller using wing-tip linear acceleration measurements. In~\cite{Baldelli2008a}, an aeroelastic morphing vehicle is controlled using linear parameter-varying (LPV) and pole placement techniques.

In~Ref.~\cite{Nguyen2017}, a distributed control surface layout named variable camber continuous trailing edge flap (VCCTEF) is used to achieve multi-objective flight control and maneuver load alleviation. Two cost functions are used in the LQG control, one for rigid-body command tracking, and another for elastic mode suppression and wing root bending moment minimization~\cite{Nguyen2017}. Simulation results in~\cite{Nguyen2017} show that the pitch rate tracking performance is degraded by the MLA function. However, for an aircraft with distributed wing control surfaces, it is physically realistic to simultaneously achieve the necessary loads for command tracking, while alleviating the excessive loads caused by maneuvers and gusts. The LQG controller in~\cite{Nguyen2017} is augmented with an adaptive GLA function in~\cite{Nguyen2018d}. Based on the differences between the measured and model-predicted accelerations, the gust components on rigid-body and elastic dynamics are estimated online using a gradient adaptive law. However, because the number of accelerometer outputs is less than the number of gust load elements, the gust estimation is only in a least-squares sense. Moreover, the resulting estimations are not purely gusts, but also contain model uncertainties. Furthermore, as commented in~\cite{Lee2018a}, atmospheric disturbances have high-frequency components, which would require prohibitively high learning rates of adaptation laws. 

Different from the above-mentioned linear model-based control techniques, an incremental nonlinear dynamic inversion (INDI) control law is proposed in~\cite{Wang2019c} for alleviating the gust loads of a flexible aircraft. In contrast to the linear control methods, INDI does not need the tedious gain-scheduling process when applied to nonlinear dynamic systems. In comparison to other model-based nonlinear control methods such as feedback linearization~\cite{Tekin2016} and backstepping~\cite{Kalliny2018a}, INDI has less model dependency, which simplifies its implementation process. Although its model dependency is reduced, the robustness of INDI is actually enhanced by exploiting the sensor measurements. Experimental and simulation results have demonstrated the robustness of INDI to model uncertainties~\cite{Grondman2018}, gust disturbances~\cite{Wang2019c}, actuator faults~\cite{Sun2020}, and structural damage~\cite{Wang2018b}. As oppose to adaptive control methods, INDI does not need the uncertainty parameterization process nor the assumption of slowly time-varying uncertain parameters. Additionally, its computational load is also lower than the adaptive control methods. 

The flexible aircraft configuration used in~\cite{Wang2019c} only has one aileron on each wing. Consequently, within the INDI control loop, trade-offs among different virtual control components have to be made. Besides, input constraints are not considered in~\cite{Wang2019c}. For the SmartX-Alpha morphing wing with distributed actuation, wing load alleviation becomes an over-actuated problem, where control allocation is needed. Moreover, it is crucial to constrain the relative deflections between adjacent morphing modules to avoid over-stretching the elastomer. In the literature, an incremental nonlinear control allocation (INCA) method has been proposed for a tailless aircraft with innovative control effectors (ICE)~\cite{Matamoros2018a}. However, its derivations based on the time-scale separation principle is not rigorous. Moreover, the relative position constraints are also not considered in~\cite{Matamoros2018a}. Furthermore, the closed-loop stability in the presence of model uncertainties, disturbances, and possible control allocation errors has not been addressed. 

The theoretical contributions of this paper are: 1) derivation and Lyapunov-based stability analysis for INDI control under the perturbation of control allocation errors, model uncertainties, and external disturbances; 2) proposal of INDI with quadratic programming control allocation (INDI-QP) considering actuator \textit{relative} position constraints, position constraints, and rate constraints; 3) proposal of INDI-QP augmented with virtual shape functions (denoted as INDI-QP-V), which can ensure the smoothness of a morphing wing at every moment. 

The practical contributions of this paper are: 1) implementation and wind tunnel testing of INDI-QP-V on a simultaneous gust and maneuver load alleviation problem of a seamless active morphing wing; 2) robust load alleviation performance comparisons between INDI-QP-V and LQG control in the presence of actuator fault and nonlinear backlash. 

The rest of this paper is structured as follows. Section~\ref{sec_control_algorithms} derives the control algorithms. The wind tunnel experiment results are presented in Sec.~\ref{sec_experiments}. The proposed INDI-QP-V control method is compared to the LQG control in Sec.~\ref{sec_LQG}. Main conclusions are drawn in Sec.~\ref{sec_conclusions}.

\section{Incremental Control Design}
\label{sec_control_algorithms}
\subsection{Incremental Control Theory}
\label{sec_sub_theory}
Consider a multi-input/multi-output nonlinear system described by 
\begin{equation}
\dot {\boldsymbol x} = \boldsymbol f(\boldsymbol x) +\boldsymbol G(\boldsymbol x) \boldsymbol u  +  \boldsymbol d(t),~~\boldsymbol y = \boldsymbol h(\boldsymbol x)
\label{non_sys}
\end{equation}
where $\boldsymbol f: \mathbb{R}^n\rightarrow \mathbb{R}^n$ and $\boldsymbol h : \mathbb{R}^n  \rightarrow \mathbb{R}^p$ are smooth vector fields. $\boldsymbol G $ is a smooth function mapping $\mathbb{R}^n \rightarrow \mathbb{R}^{n\times m}$, whose columns are smooth vector fields. $\boldsymbol d(t) \in \mathbb{R}^n$ represents the external disturbance vector. Assume $\|\boldsymbol d(t)\|_2 \leq \bar d$. $\boldsymbol y \in \mathbb{R}^p$ in Eq.~(\ref{non_sys}) denotes the controlled output vector, which can be a function of any subset of the physical measurable outputs. This paper considers the case where $p\leq m$. Define the vector relative degree~\cite{Fradkov} of the system as $\boldsymbol \rho=  [\rho_1,\rho_2,...,\rho_p]^\mathsf{T}$, which satisfies $\rho = \|\boldsymbol \rho\|_1 = \sum_{i = 1}^p \rho_i \leq n$, then by differentiating the output vector $\boldsymbol y$, the input–output mapping is given as
\begin{equation}
\boldsymbol y^{(\boldsymbol \rho)} = \boldsymbol \alpha (\boldsymbol x) + \boldsymbol{\mathcal{B}}(\boldsymbol x) \boldsymbol u + \boldsymbol d_y
\label{y_rho}
\end{equation}

In Eq.~(\ref{y_rho}), $\boldsymbol \alpha (\boldsymbol x) = [\mathcal{L}_f^{\rho_1}h_1, \mathcal{L}_f^{\rho_2}h_2,...,\mathcal{L}_f^{\rho_p}h_p ]^\mathsf{T}$, $ \boldsymbol{\mathcal{B}}(\boldsymbol x) \in \mathbb{R}^{p \times m}$, $\mathcal{B}_{ij} = \mathcal{L}_{g_j}\mathcal{L}_{f}^{\rho_i-1}h_i$, where $\mathcal{L}_f^{\rho_i} h_i,~\mathcal{L}_{ g_j} \mathcal{L}_f^{\rho_i-1} h_i$ are the corresponding Lie derivatives~\cite{Khalil}. When $\rho_i =1$ for all $~i=1,...,p$, $\boldsymbol d_y = [\mathcal{L}_d h_1, \mathcal{L}_d h_2,...,\mathcal{L}_d h_p ]^\mathsf{T}$. For more general cases where $\rho_i>1$, $\boldsymbol d_y$ also contains the cross-coupling terms of $\mathcal{L}_d h_i $ and $\mathcal{L}_f h_i$. If $\rho = n$, then the system given by Eq.~(\ref{non_sys}) is full-state feedback linearizable. Otherwise, there exists $n-\rho$ internal dynamics. 

Denote the sampling interval as $\Delta t$, the incremental dynamic equation is derived by taking the first-order Taylor series expansion of Eq.~(\ref{y_rho}) around the condition at $t-\Delta t$ (denoted by the subscript 0) as:
\begin{equation}
\boldsymbol y^{(\boldsymbol \rho)} =
\boldsymbol y^{(\boldsymbol \rho)}_0 + \frac{\partial [\boldsymbol \alpha(\boldsymbol x) + \boldsymbol{\mathcal{B}}(\boldsymbol x)\boldsymbol u]}{\partial \boldsymbol x}\bigg|_0 \Delta \boldsymbol x + \boldsymbol{\mathcal{B}}(\boldsymbol x_0)\Delta \boldsymbol u + \Delta \boldsymbol d_y + \boldsymbol{R}_1
\label{chp3_y_rho_expand}
\end{equation}
in which $\Delta \boldsymbol x$, $\Delta \boldsymbol  u$, and $\Delta \boldsymbol d_y $ respectively represents the state, control, and disturbance increments in one sampling time step $\Delta t $. $\boldsymbol{R}_1$ is the expansion remainder. Consider the output tracking problem, and denote the output reference signal as $\boldsymbol y_{r}(t) = [y_{r_1}(t),y_{r_2}(t),...,y_{r_p}(t)]^\mathsf{T}$. Assume $y_{r_i}(t),~i=1,2,...,p$, and its derivatives up to $y_{r_i}^{(\rho_i)}(t)$ are bounded for all $t$ and each $y_{r_i}^{(\rho_i)}(t)$ is continuous, then the tracking error vector yields $\boldsymbol{e} =  \boldsymbol \xi -\boldsymbol {\mathcal R}, ~
\boldsymbol {\mathcal R} = [\boldsymbol {\mathcal R}_1^\mathsf{T},\boldsymbol {\mathcal R}_2^\mathsf{T},...,\boldsymbol {\mathcal R}_p^\mathsf{T}]^\mathsf{T}, ~ \boldsymbol {\mathcal R}_i = [y_{r_i},y_{r_i}^{(1)},...,y_{r_i}^{(\rho_i-1)}]^\mathsf{T}$. Assume $\|\boldsymbol {\mathcal R} \|_2 \leq \bar {\mathcal R} $. To stabilize the error dynamics, the control increment is designed to satisfy the following equation:
\begin{equation}
\bar{ {\boldsymbol{\mathcal{B}}}}(\boldsymbol x_0) \Delta \boldsymbol u_{\text{indi}} = \boldsymbol \nu_c -\boldsymbol y^{(\boldsymbol \rho)}_0 ,~~~~\boldsymbol \nu_c =  \boldsymbol y_r^{(\boldsymbol \rho)} - \boldsymbol K \boldsymbol e
\label{u_indi_track}
\end{equation}
where $\bar{ {\boldsymbol{\mathcal{B}}}}$ is an estimation of $ {\boldsymbol{\mathcal{B}}}$. The gain matrix $\boldsymbol K = \text{diag}\{\boldsymbol K_i\},~i= 1,2,...,p$, and $\boldsymbol K_i = [K_{i,0},K_{i,1},...,K_{i,\rho_i-1}]$. $ \boldsymbol y^{(\boldsymbol \rho)}_0 $ is directly measured or estimated. The total control command for actuator is $\boldsymbol u_{\text{indi}} = \boldsymbol u_{\text{indi},0} + \Delta \boldsymbol u_{\text{indi}}$. Assume the roll rank of $\bar{ {\boldsymbol{\mathcal{B}}}}$ equals $p$. If the column rank of $\bar{ {\boldsymbol{\mathcal{B}}}}$ also equals $p$, then there exists a unique $\Delta \boldsymbol u_{\text{indi}}$ satisfying Eq.~\eqref{u_indi_track}. If the column rank of $\bar{ {\boldsymbol{\mathcal{B}}}}$ is less than $p$, then the system is under-actuated and Eq.~(\ref{u_indi_track}) cannot be satisfied. If the column rank of $\bar{ {\boldsymbol{\mathcal{B}}}}$ is larger than $p$, then solving $\Delta \boldsymbol u_{\text{indi}}$ from Eq.~\eqref{u_indi_track} is a control allocation problem. Without considering the input constraints of $\boldsymbol u_{\text{indi}}$, there are infinite $\Delta \boldsymbol u_{\text{indi}}$ that satisfies Eq.~\eqref{u_indi_track}. However, when some dimensions of $\boldsymbol u_{\text{indi}}$ get saturated, it is possible that $\bar{ {\boldsymbol{\mathcal{B}}}}(\boldsymbol x_0) \Delta \boldsymbol u_{\text{indi}} - (\boldsymbol \nu_c -\boldsymbol y^{(\boldsymbol \rho)}_0) \neq \boldsymbol{0}$ \textit{even though} the column rank of $\bar{ {\boldsymbol{\mathcal{B}}}}$ is higher than $p$. To keep the theoretical analyses more general, Eq.~\eqref{u_indi_track} is generalized to $\bar{ {\boldsymbol{\mathcal{B}}}}(\boldsymbol x_0) \Delta \boldsymbol u_{\text{indi}}  =  \boldsymbol \nu_c -\boldsymbol y^{(\boldsymbol \rho)}_0 + \boldsymbol \varepsilon_{\text{ca}}$, with $\boldsymbol \varepsilon_{\text{ca}}$ indicates the possible control allocation error. Considering the internal dynamics, the resulting closed-loop dynamics are:
\begin{eqnarray}
\dot {\boldsymbol \eta} &=& \boldsymbol f_\eta (\boldsymbol \eta,\boldsymbol \xi,\boldsymbol d) = \frac{\partial \boldsymbol \phi}{\partial \boldsymbol x} (\boldsymbol f(\boldsymbol x)  +\boldsymbol{d}(t))\bigg|_{\boldsymbol x =\boldsymbol T^{-1}(\boldsymbol z)}
\nonumber \\
\dot{\boldsymbol e} &=&  (\boldsymbol A_c -\boldsymbol B_c \boldsymbol K) \boldsymbol e + \boldsymbol B_c[\boldsymbol \delta(\boldsymbol x, \Delta t) + (\boldsymbol{\mathcal{B}}(\boldsymbol x_0) - \bar{ {\boldsymbol{\mathcal{B}}}}(\boldsymbol x_0)) \Delta \boldsymbol u_{\text{indi}} + 
\boldsymbol \varepsilon_{\text{ca}} + \Delta \boldsymbol d_y ]\nonumber \\
 &\triangleq& (\boldsymbol A_c -\boldsymbol B_c \boldsymbol K) \boldsymbol e + \boldsymbol B_c \boldsymbol \varepsilon_{\text{indi}}
\label{closed_loop_indi}
\end{eqnarray}
where $\boldsymbol \eta $ represents the internal state vector, and $\boldsymbol z=\boldsymbol T(\boldsymbol x) =  [\boldsymbol \eta^\mathsf{T},\boldsymbol \xi^\mathsf{T}]^\mathsf{T}$, $\boldsymbol \eta = \boldsymbol \phi (\boldsymbol x)$, $\boldsymbol \xi = [\boldsymbol \xi_1^\mathsf{T},...,\boldsymbol \xi_p^\mathsf{T}]^\mathsf{T} $, $\boldsymbol \xi_i = [h_i(\boldsymbol x),...,\mathcal{L}_f^{\rho_i-1} h_i(\boldsymbol x)]^\mathsf{T}$, $ i=1,2,...,p $ is a diffeomorphism. $\boldsymbol \delta(\boldsymbol x, \Delta t) $ is the closed-loop value of the variations and expansion reminder: $   \boldsymbol \delta(\boldsymbol x, \Delta t) = \left[ \frac{\partial [\boldsymbol \alpha(\boldsymbol x) + \boldsymbol{\mathcal{B}}(\boldsymbol x)\boldsymbol u]}{\partial \boldsymbol x}\big|_0 \Delta \boldsymbol x  + \boldsymbol{R}_1 \right]\Big|_{\boldsymbol{u} = \boldsymbol u_{\text{indi}}}$. $\boldsymbol A_c = \text{diag}\{\boldsymbol A_0^i\}$, $\boldsymbol B_c = \text{diag}\{\boldsymbol B_0^i\}$, $\boldsymbol C_c = \text{diag}\{\boldsymbol C_0^i\}$, $i= 1,2,...,p$, and $(\boldsymbol A_0^i,\boldsymbol B_0^i,\boldsymbol C_0^i)$ is a canonical form representation of a chain of $\rho_i$ integrators. The gain matrix $\boldsymbol K$ is designed such that $\boldsymbol A_c - \boldsymbol B_c \boldsymbol K$ is Hurwitz.

In contrast to the model-based feedback linearization, the incremental nonlinear dynamic inversion (INDI) is a sensor-based control strategy~\cite{Wang2019b}. By exploiting the sensor measurements, the only model information needed by INDI is the estimated control effectiveness matrix $\bar{ {\boldsymbol{\mathcal{B}}}}$, which simplifies the implementation process. Moreover, the residual perturbation in the closed-loop system is also reduced, which enhances the control robustness against model uncertainties, external disturbances, and sudden faults~\cite{Wang2018b}.

\begin{remark}
\rm A stability analysis for INDI that simultaneously considers control allocation errors, internal dynamics, model uncertainties, and external disturbances has not been addressed in the literature. In view of this, the following two theorems are proposed in this paper:
\end{remark}
\begin{theorem}
\label{theorem_1}
If $ \|\boldsymbol \varepsilon_{\text{indi}} \|_2 \leq \bar{ \varepsilon}$ is satisfied for all $\boldsymbol \xi \in \mathbb{R}^{\rho} $, $ \boldsymbol f_\eta (\boldsymbol \eta,\boldsymbol \xi,\boldsymbol d)$ is continuously differentiable and globally Lipschitz in $(\boldsymbol \eta,\boldsymbol \xi,\boldsymbol d)$, and the origin of $\dot {\boldsymbol \eta } =  \boldsymbol f_\eta (\boldsymbol \eta,\boldsymbol 0,\boldsymbol 0)$ is globally exponentially stable, then the tracking error $\boldsymbol e$ in Eq.~\eqref{closed_loop_indi} is globally ultimately bounded by a class $\mathcal K$ function of $\bar \varepsilon$, while the internal state $\boldsymbol \eta$ in Eq.~\eqref{closed_loop_indi} is globally ultimately bounded by a class $\mathcal{K}$ function of $\bar \varepsilon$, $\bar{\mathcal R}$, and $\bar d$.
\end{theorem}

\noindent\textit{\textbf{Proof}}: See Appendix.

\begin{theorem}
\label{theorem_2}
If $ \|\boldsymbol \varepsilon_{\text{indi}} \|_2 \leq \bar{ \varepsilon}$ is satisfied for all $\boldsymbol \xi \in \mathbb{R}^{\rho} $, $ \boldsymbol f_\eta (\boldsymbol \eta,\boldsymbol \xi,\boldsymbol d)$ is continuously differentiable, and the origin of $\dot {\boldsymbol \eta } =  \boldsymbol f_\eta(\boldsymbol \eta,\boldsymbol 0,\boldsymbol 0)$ is exponentially stable, then there exists a neighborhood $D_z$ of $\boldsymbol z =[\boldsymbol 0^\mathsf{T}, \boldsymbol {\mathcal R}^\mathsf{T}]^\mathsf{T}$ and $\varepsilon^*>0$, such that for every $\boldsymbol z(t=0)\in D_z$ and $\bar {\varepsilon}<\varepsilon^*$, the tracking error $\boldsymbol e$ in Eq.~\eqref{closed_loop_indi} is ultimately bounded by a class $\mathcal{K}$ function of $\bar \varepsilon$, while the internal state $\boldsymbol \eta$ in Eq.~\eqref{closed_loop_indi} is ultimately bounded by a class $\mathcal{K}$ function of $\bar \varepsilon$, $\bar{\mathcal R}$, and $\bar d$.
\end{theorem}

\noindent\textit{\textbf{Proof}}: See Appendix. 


\subsection{Incremental Control Allocation}
\label{sec_sub_CA}

This subsection will solve $\Delta \boldsymbol u_{\text{indi}}$ from Eq.~\eqref{u_indi_track}, and discuss the corresponding boundedness conditions for $\boldsymbol \varepsilon_{\text{indi}}$ (Eq.~\eqref{closed_loop_indi}). The control allocation problem considers the case that the roll rank of $ \bar{\boldsymbol{\mathcal{B}}} \in \mathbb{R}^{p \times m}$ equals $p$, while its column rank is larger than $p$. Under this condition, Eq.~\eqref{u_indi_track} is satisfied by $ \Delta \boldsymbol u_{\text{indi}} = \bar{\boldsymbol{\mathcal{B}}}^{+}(\boldsymbol x_0) (\boldsymbol \nu_c -\boldsymbol y^{(\boldsymbol \rho)}_0 )+ (\boldsymbol{I}_{m\times m} -  \bar{\boldsymbol{\mathcal{B}}}^{+}(\boldsymbol x_0) \bar{\boldsymbol{\mathcal{B}}}(\boldsymbol x_0) ) \boldsymbol{w}$. In this equation, $\bar{\boldsymbol{\mathcal{B}}}^{+} = \bar{\boldsymbol{\mathcal{B}}}^{T}(\bar{\boldsymbol{\mathcal{B}}}\bar{\boldsymbol{\mathcal{B}}}^{T})^{-1}$ is the Moore-Penrose inverse of $\bar{\boldsymbol{\mathcal{B}}}$. It is noteworthy that although $\bar{\boldsymbol{\mathcal{B}}}\bar{\boldsymbol{\mathcal{B}}}^{+} = \boldsymbol{I}_{p\times p}$, $\bar{\boldsymbol{\mathcal{B}}}^{+}\bar{\boldsymbol{\mathcal{B}}} \neq \boldsymbol{I}_{m\times m}$. Besides, $\boldsymbol{w}$ can be any vector in $\mathbb{R}^{m \times 1}$. Nevertheless, $\Delta \boldsymbol u_{\text{indi}}$ only has the smallest Euclidean norm when $\boldsymbol{w} = \boldsymbol{0}$. This least squares solution given by pseudo-inverse is:
\begin{equation}
    \Delta \boldsymbol u_{\text{indi-pi}} = \bar{\boldsymbol{\mathcal{B}}}^{+}(\boldsymbol x_0) (\boldsymbol \nu_c -\boldsymbol y^{(\boldsymbol \rho)}_0 )
     \label{u_indi_pi}
\end{equation}

\begin{theorem}
\label{theorem_3}
When the pseudo-inverse control allocation is used (Eq.~(\ref{u_indi_pi})), if $\|\boldsymbol I-  {\boldsymbol{\mathcal{B}}}(\boldsymbol x_0)  \bar{\boldsymbol{\mathcal{B}}}^{+}(\boldsymbol x_0) \|_2 \leq \bar b <1$, and if $\boldsymbol \delta(\boldsymbol x, \Delta t) $ and $\Delta \boldsymbol{d}_y$ are respectively bounded by $\bar{\delta}$ and $\overline{\Delta d}$, then under sufficiently high sampling frequency, $\boldsymbol \varepsilon_{\text{indi}}$ in Eq.~(\ref{closed_loop_indi}) is ultimately bounded.
\end{theorem}

\noindent\textit{\textbf{Proof}}: See Appendix. 

Theorem~\ref{theorem_3} presents that one of the sufficient conditions for the boundedness of $\boldsymbol \varepsilon_{\text{indi}}$ is a diagonally dominated ${\boldsymbol{\mathcal{B}}}(\boldsymbol x_0)  \bar{\boldsymbol{\mathcal{B}}}^{+}(\boldsymbol x_0)$. If this condition is satisfied, then the influences of model mismatches can be automatically tolerated by the controller. Otherwise, online model identification and adaptation for $\bar{\boldsymbol{\mathcal{B}}}(\boldsymbol x_0)$ can be needed. 

Although pseudo-inverse can provide the least squares solution, the input constraints are not considered. The servo position constraints are formulated as $\boldsymbol{ u}_{\text{min}} \leq \boldsymbol{ u} \leq \boldsymbol{ u}_{\text{max}}$, which can be rewritten as a linear inequality $[\boldsymbol{I}_{m\times m},-\boldsymbol{I}_{m\times m}]^\mathsf{T} (\Delta \boldsymbol{ u}+ \boldsymbol{ u}_0 ) \leq [\boldsymbol{ u}^\mathsf{T}_{\text{max}}, - \boldsymbol{ u}^\mathsf{T}_{\text{min}}]^\mathsf{T}$. The servos also have rate limits, i.e., $ \underline{\boldsymbol{ u}}_{\text{rate}}\Delta t \leq \Delta \boldsymbol{ u} \leq \bar{\boldsymbol{ u}}_{\text{rate}}\Delta t$. Furthermore, to avoid the elastomer between the morphing modules being over-stretched, the relative command differences between adjacent servos also need to be constrained. For $ \boldsymbol{ u}\in \mathbb{R}^{m}$, there are $m-1$ relative position constraints. Denote them as $ \bar{\boldsymbol{ u}}_{\text{adj}}\in \mathbb{R}^{m-1}$, with $|u_{i+1} - u_i | \leq  \bar{ u}_{\text{adj},i},~i =1,2,...,m-1$. The elements of $ \bar{\boldsymbol{ u}}_{\text{adj}}$ are not necessarily equal. For example, regarding two adjacent servos in the SmartX-Alpha wing, if it is elastomer between them, then the relative actuation limit is set as $10$~deg to prevent over-stretching. Otherwise, the relative actuation limit is relaxed to $55$~deg. The relative position constrains are formulated as the following inequality: $[\boldsymbol{C},-\boldsymbol{C}]^\mathsf{T} (\Delta \boldsymbol{ u}+ \boldsymbol{ u}_0 ) \leq [\bar{\boldsymbol{ u}}^\mathsf{T}_{\text{adj}},  \bar{\boldsymbol{ u}}^\mathsf{T}_{\text{adj}}]^\mathsf{T}$. $\boldsymbol{C} \in \mathbb{R}^{(m-1)\times m}$, with $C_{i,i} = 1$ and $C_{i,i+1} = -1$ for $i = 1,2,...,m-1$. Besides, the rest elements of $\boldsymbol{C}$ are all equal to zero. Considering the servo position, rate, and relative position limits, the control increment vector $\Delta \boldsymbol{u}$ has to satisfy the following inequality:
\begin{spacing}{1.3}
\begin{eqnarray}
\begin{bmatrix}
\boldsymbol{I}_{m\times m} \\ 
-\boldsymbol{I}_{m\times m} \\ \hdashline
\boldsymbol{C} \\
-\boldsymbol{C} \\ \hdashline
\boldsymbol{I}_{m\times m} \\
-\boldsymbol{I}_{m\times m} 
\end{bmatrix}
\Delta \boldsymbol{u} \leq
\begin{pmatrix}
\boldsymbol{ u}_{\text{max}} -  \boldsymbol{u}_0 \\
- \boldsymbol{ u}_{\text{min}} +  \boldsymbol{u}_0 \\ \hdashline
\bar{\boldsymbol{ u}}_{\text{adj}} - \boldsymbol{C} \boldsymbol{u}_0 \\
\bar{\boldsymbol{ u}}_{\text{adj}} + \boldsymbol{C} \boldsymbol{u}_0  \\ \hdashline
\bar{\boldsymbol{ u}}_{\text{rate}}\Delta t \\
-\underline{\boldsymbol{ u}}_{\text{rate}}\Delta t 
\end{pmatrix},~~\text{denoted as}~~ \boldsymbol A_u\Delta \boldsymbol{u} \leq \boldsymbol b_u 
\label{inequality}
\end{eqnarray}
\end{spacing}

\begin{remark}
\rm Equation~\eqref{inequality} presents the first work that converts the actuator position constraints, rate constraints, and \textit{relative} position constraints into an integrated linear inequality matrix with respect to the incremental control vector $\Delta \boldsymbol{u}$. 
\end{remark}

From a theoretical point of view, the linear equality constraint in Eq.~\eqref{u_indi_track} has the highest priority. If both Eq.~\eqref{u_indi_track} and the inequality constraint in Eq.~\eqref{inequality} can be satisfied, then the rest free space of $\Delta \boldsymbol{u}$ can be used to minimize the energy of $\boldsymbol{u}$. However, under some faulty conditions, the feasible region can become null if both the equality (Eq.~\eqref{u_indi_track}) and the inequality (Eq.~\eqref{inequality}) constraints are imposed. Actually, it is more practical to satisfy the inequality first, and then minimize the realization error of the equality constraint. For example, consider an actuator fault condition where Eq.~\eqref{u_indi_track} and Eq.~\eqref{inequality} cannot be simultaneously satisfied; it is more meaningful to realize Eq.~\eqref{inequality} first and allow certain performance degradation, rather than enforcing Eq.~\eqref{u_indi_track} by violating Eq.~\eqref{inequality}. Therefore, the first cost function is formulated as $\mathcal{J}_1 =(1/2)(\bar{ {\boldsymbol{\mathcal{B}}}}(\boldsymbol x_0) \Delta \boldsymbol u  - \boldsymbol \nu_c +\boldsymbol y^{(\boldsymbol \rho)}_0)^\mathsf{T} \boldsymbol{W}_1 (\bar{ {\boldsymbol{\mathcal{B}}}}(\boldsymbol x_0) \Delta \boldsymbol u  - \boldsymbol \nu_c +\boldsymbol y^{(\boldsymbol \rho)}_0) $, where $\boldsymbol{W}_1$ is a positive definite weighting matrix. 

Apart from realizing Eq.~\eqref{u_indi_track}, the control allocator should make $\boldsymbol{u}$ close to its nominal value $\boldsymbol{u}_*$. A typical choice is $\boldsymbol{u}_* = \boldsymbol{0}$ for minimizing the control energy. For a morphing wing, $\boldsymbol{u}_*$ can also be non-zero to achieve an optimized wing shape. Therefore, the second cost function is $\mathcal{J}_2 = (1/2)(\Delta \boldsymbol u +\boldsymbol{u}_0 - \boldsymbol{u}_*)^\mathsf{T} \boldsymbol{W}_2 (\Delta \boldsymbol u +\boldsymbol{u}_0 - \boldsymbol{u}_*)$, where $\boldsymbol{W}_2$ is another positive definite weighting matrix. Choose $\mathcal{J}_3 = \mathcal{J}_1 + \sigma\mathcal{J}_2$, where $0<\sigma \ll 1$ for prioritizing $\mathcal{J}_1$. Further derive $\mathcal{J}_3 $ as
\begin{eqnarray}
\mathcal{J}_3 = \mathcal{J}_1 + \sigma\mathcal{J}_2 &= & \frac{1}{2} \Delta \boldsymbol u^\mathsf{T}\left( \bar{ {\boldsymbol{\mathcal{B}}}}^\mathsf{T}(\boldsymbol x_0) \boldsymbol{W}_1\bar{ {\boldsymbol{\mathcal{B}}}}(\boldsymbol x_0) + \sigma \boldsymbol{W}_2 \right)\Delta \boldsymbol u + \left((\boldsymbol y^{(\boldsymbol \rho)}_0- \boldsymbol \nu_c )^\mathsf{T} \boldsymbol{W}_1\bar{ {\boldsymbol{\mathcal{B}}}}(\boldsymbol x_0) + (\boldsymbol{u}_0 - \boldsymbol{u}_*)^\mathsf{T} \sigma \boldsymbol{W}_2   \right) \Delta \boldsymbol{u} \nonumber \\
&&+ \frac{1}{2}\left( (\boldsymbol y^{(\boldsymbol \rho)}_0- \boldsymbol \nu_c )^\mathsf{T} \boldsymbol{W}_1 (\boldsymbol y^{(\boldsymbol \rho)}_0- \boldsymbol \nu_c )+ (\boldsymbol{u}_0 - \boldsymbol{u}_*)^\mathsf{T} \sigma \boldsymbol{W}_2 (\boldsymbol{u}_0 - \boldsymbol{u}_*)\right)
\end{eqnarray}

Since within every time step, $\boldsymbol{u}_0$ and $\boldsymbol y^{(\boldsymbol \rho)}_0$ are measured, while $\boldsymbol{u}_*$ and $\boldsymbol \nu_c$ are constants, only the terms related to $\Delta \boldsymbol{u}$ need to be minimized. Therefore, the incremental control allocation problem is formulated as:
\begin{eqnarray}
\mathop{\text{min}}\limits_{ \Delta \boldsymbol{u}} \mathcal{J}_4 &= &\frac{1}{2} \Delta \boldsymbol u^\mathsf{T}\left( \bar{ {\boldsymbol{\mathcal{B}}}}^\mathsf{T}(\boldsymbol x_0) \boldsymbol{W}_1\bar{ {\boldsymbol{\mathcal{B}}}}(\boldsymbol x_0) + \sigma \boldsymbol{W}_2 \right)\Delta \boldsymbol u + \left((\boldsymbol y^{(\boldsymbol \rho)}_0- \boldsymbol \nu_c )^\mathsf{T} \boldsymbol{W}_1\bar{ {\boldsymbol{\mathcal{B}}}}(\boldsymbol x_0) + (\boldsymbol{u}_0 - \boldsymbol{u}_*)^\mathsf{T} \sigma \boldsymbol{W}_2   \right) \Delta \boldsymbol{u}, \nonumber \\  
&&  \text{subject to}~~ \boldsymbol A_u\Delta \boldsymbol{u} \leq \boldsymbol b_u 
\label{QP_J}
\end{eqnarray}

The active-set solver is selected because of its superior performance on solving small to medium size quadratic programming problems~\cite{bartlett2000active}. In contrast to the $\Delta \boldsymbol u_{\text{indi-pi}}$ in Eq.~\eqref{u_indi_pi}, it is difficult to write an analytical expression for the control input given by quadratic programming. Consequently, Theorem~\ref{theorem_3} is not applicable here. To derive a sufficient condition for the boundedness of $\boldsymbol \varepsilon_{\text{indi}}$, when the quadratic programming allocator is applied, assume at every time step, $ {\boldsymbol{\mathcal{B}}}(\boldsymbol x_0) = \boldsymbol{K}_{\boldsymbol{\mathcal{B}}}(\boldsymbol x_0)\bar{ {\boldsymbol{\mathcal{B}}}}(\boldsymbol x_0)$, then the following theorem holds:

\begin{theorem}
\label{theorem_4}
When the quadratic programming control allocation is used (Eq.~(\ref{QP_J})), if $\|\boldsymbol I-   \boldsymbol{K}_{\boldsymbol{\mathcal{B}}} (\boldsymbol{x}_0)\|_2 \leq \bar {b}' <1$, and if $\boldsymbol \delta(\boldsymbol x, \Delta t) $, $\Delta \boldsymbol{d}_y$, and $\boldsymbol \varepsilon_{\text{ca}}$ are respectively bounded by $\bar{\delta}$, $\overline{\Delta d}$, and $\bar{ \varepsilon}_{\text{ca}}$, then under sufficiently high sampling frequency, $\boldsymbol \varepsilon_{\text{indi}}$ in Eq.~(\ref{closed_loop_indi}) is ultimately bounded.
\end{theorem}

\noindent\textit{\textbf{Proof}}: See Appendix.


\subsection{Virtual Shape Functions}
\label{sec_sub_VC}

In the preceding subsections, the number of control input equals the number of servos. Although the relative command differences of any adjacent servos have been constrained by Eq.~\eqref{inequality}, the resulting $\boldsymbol{u} \in \mathbb{R}^{m \times 1}$ does not necessarily lead to a smooth wing shape. This subsection will introduce virtual shape functions to solve this problem. 

Define a reference axis where $O_s$ is located at the wing root, while $O_s x_s$ is aligned with the servo line (Fig.~\ref{fig_morphing_mechanism}). The aim is to make the morphing wing trailing-edge shape as close as possible to a smooth function $f_s(t,x_s): [0,\infty)\times \mathbb{R} \rightarrow \mathbb{R}$. Referring to the Weierstrass theorem, when $q$ is sufficiently large, any sufficiently smooth function can be approximated by a $q$-th order polynomial, i.e., $f_s(x_s,t) \approx \tilde f_s(x_s,t) =  \boldsymbol{\Theta}^\mathsf{T}(t) \boldsymbol\Phi({x_s})$, with $\boldsymbol{\Theta}(t):[0,\infty)\rightarrow \mathbb{R}^{q\times 1},~\boldsymbol\Phi({x_s}):\mathbb{R}\rightarrow \mathbb{R}^{q\times 1} $. The Chebyshev polynomials are selected in this paper because of their nearly optimal property and orthogonality~\cite{Gomroki2018}. Design the virtual shape function as $ \boldsymbol\Phi({x_s}) = [T_0(x_s),T_1(x_s),...,T_q(x_s)]^\mathsf{T}$, whose elements are the Chebyshev polynomials of the first kind: $ T_1(x_s) = 1, ~T_2(x_s) = x_s,~T_{i+1} = 2x_s T_i (x_s) -T_{i-1}(x_s) , ~i = 2,3,...,q-1 $. Consequently, any $\boldsymbol{\Theta}(t) = [\theta_1(t), \theta_2(t),...,\theta_q(t)]^\mathsf{T}$ guarantees the $q$-th order smoothness of $\tilde f_s(t,x_s)$. Denote the servo spanwise location vector as $\boldsymbol{x}_s = [x_{s,1},x_{s,2},...,x_{s,m}]^\mathsf{T}$, which can be normalized by the half-wing span $L$ as $\bar{\boldsymbol{x}}_s = [x_{s,1}/L,x_{s,2}/L,...,x_{s,m}/L]^\mathsf{T}$. Substitute the normalized servo location vector into $\tilde f_s(x_s,t)$ yields
\begin{spacing}{1.3}
\begin{eqnarray}
\begin{pmatrix}
\tilde f_s(\bar x_{s,1},t) \\
\tilde f_s(\bar x_{s,2},t) \\
\vdots \\
\tilde f_s(\bar x_{s,m},t) \\
\end{pmatrix}
= 
\begin{bmatrix}
T_0(\bar x_{s,1}) & T_1(\bar x_{s,1})& ... &T_q(\bar x_{s,1}) \\
T_0(\bar x_{s,2})&T_1(\bar x_{s,2})&...&T_q(\bar x_{s,2}) \\
\vdots & \vdots &  & \vdots \\
T_0(\bar x_{s,m}) & T_1(\bar x_{s,m})& ... &T_q(\bar x_{s,m}) \\
\end{bmatrix}
\begin{pmatrix}
\theta_1(t)\\
\theta_2(t)\\
\vdots \\
\theta_q(t)
\end{pmatrix}
\triangleq \boldsymbol {\Phi}_{\bar{x}_s} \boldsymbol{\Theta} (t)
\label{virtual_shape}
\end{eqnarray}
\end{spacing}
\noindent where $\boldsymbol {\Phi}_{\bar{x}_s} \in \mathbb{R}^{m\times q}$ becomes a constant shape matrix. Essentially, $\boldsymbol {\Phi}_{\bar{x}_s}$ provides a mapping between a smooth wing shape and $\boldsymbol{\Theta} (t)$. In view of this, choose a new control vector $\boldsymbol{u}_v = \boldsymbol{\Theta} (t)\in \mathbb{R}^{q\times 1}$. If the actual control command is mapped as $\boldsymbol{u} = \boldsymbol {\Phi}_{\bar{x}_s}\boldsymbol{u}_v$, then this $\boldsymbol{u} $ can result in smooth wing shapes at all $t\in[0,\infty)$. The first five normalized virtual shape functions for the SmartX-Alpha are illustrated in Fig.~\ref{virtual_shape}. 
\begin{figure}[!h]
\centering
\includegraphics[width=.45\textwidth]{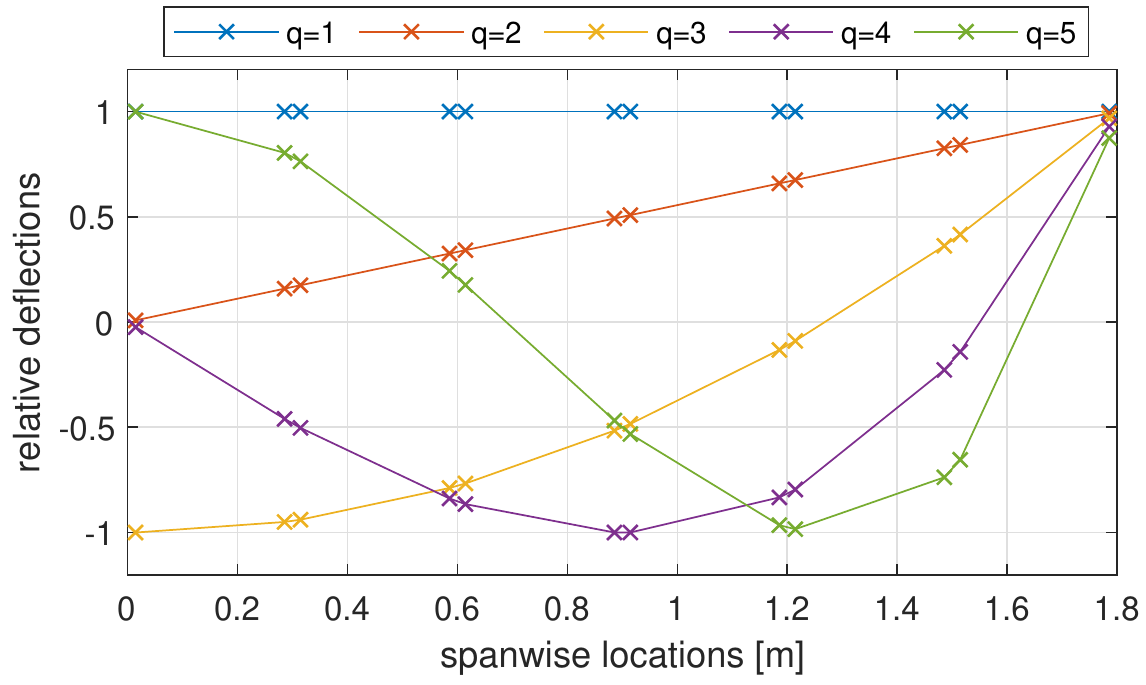}
\caption{Virtual shape functions with markers indicating the SmartX-Alpha servo locations.}
\label{virtual_shape}
\end{figure}

Because $\boldsymbol {\Phi}_{\bar{x}_s}$ is a constant matrix, this mapping also holds for the control increments, i.e., $\Delta \boldsymbol{u} = \boldsymbol {\Phi}_{\bar{x}_s} \Delta \boldsymbol{u}_v$. In essential, the control effective matrix with respect to $\Delta \boldsymbol{u}_v$ becomes $ \bar{ {\boldsymbol{\mathcal{B}}}}'(\boldsymbol x_0) = \left(\bar{ {\boldsymbol{\mathcal{B}}}}(\boldsymbol x_0) \boldsymbol {\Phi}_{\bar{x}_s} \right)\in \mathbb{R}^{p\times q}$. If the column rank of $\bar{ {\boldsymbol{\mathcal{B}}}}'$ is larger than $p$, then the quadratic programming problem integrated with virtual shape functions is formulated as
\begin{eqnarray}
\mathop{\text{min}}\limits_{ \Delta \boldsymbol{u}_v} \mathcal{J}_5 &= &\frac{1}{2} \Delta \boldsymbol u_v^\mathsf{T}\left( \boldsymbol {\Phi}^\mathsf{T}_{\bar{x}_s}\bar{ {\boldsymbol{\mathcal{B}}}}^\mathsf{T}(\boldsymbol x_0) \boldsymbol{W}_1\bar{ {\boldsymbol{\mathcal{B}}}}(\boldsymbol x_0) \boldsymbol {\Phi}_{\bar{x}_s} + \sigma \boldsymbol {\Phi}^\mathsf{T}_{\bar{x}_s}\boldsymbol{W}_2 \boldsymbol {\Phi}_{\bar{x}_s} \right)\Delta \boldsymbol u_v + \left((\boldsymbol y^{(\boldsymbol \rho)}_0- \boldsymbol \nu_c )^\mathsf{T} \boldsymbol{W}_1\bar{ {\boldsymbol{\mathcal{B}}}}(\boldsymbol x_0) + (\boldsymbol{u}_0 - \boldsymbol{u}_*)^\mathsf{T} \sigma \boldsymbol{W}_2   \right) \boldsymbol {\Phi}_{\bar{x}_s}\Delta \boldsymbol{u}_v, \nonumber \\  
&&  \text{subject to}~~ \left(\boldsymbol A_u \boldsymbol {\Phi}_{\bar{x}_s} \right)\Delta \boldsymbol{u}_v \leq \boldsymbol b_u
\label{QP_J_virtual}
\end{eqnarray}

In fact, because the dimension of $\Delta \boldsymbol{u}_v$ is lower than that of $\Delta \boldsymbol{u}$, the computational load is also reduced by introducing the virtual shape functions. In this case, the following corollary of Theorem~\ref{theorem_4} is given:
\begin{corollary}
\label{corollary_1}
When the quadratic programming control allocation with virtual shape functions is used (Eq.~(\ref{QP_J_virtual})), if $\|\boldsymbol I-   \boldsymbol{K}_{\mathcal{B}} (\boldsymbol{x}_0)\|_2 \leq \bar {b}' <1$, and if $\boldsymbol \delta(\boldsymbol x, \Delta t) $, $\Delta \boldsymbol{d}_y$, and $\boldsymbol \varepsilon_{\text{ca}}$ are respectively bounded by $\bar{\delta}$, $\overline{\Delta d}$, and $\bar{ \varepsilon}_{\text{ca}}$, then under sufficiently high sampling frequency, $\boldsymbol \varepsilon_{\text{indi}}$ in Eq.~(\ref{closed_loop_indi}) is ultimately bounded.
\end{corollary}

\noindent\textit{\textbf{Proof}}: See Appendix.

\begin{remark}
\rm The virtual shape functions were also used in~\cite{Nguyen2018d,Ferrier2018} intending to address the relative deflection constraints. However, the usage of virtual shape itself is not sufficient for meeting the relative position constraints. By contrast, the control allocator formulated in Eq.~\eqref{QP_J_virtual} not only explicitly considers the position, rate, and relative position constraints, but also leads to a smooth wing shape at every moment.
\end{remark}


\section{Experimental Results}
\label{sec_experiments}

In this section, the proposed incremental control will be applied to the SmartX-Alpha load alleviation problems. The experiment setup will be presented in Sec.~\ref{sec_sub_exp}, following which the challenges in the experiment will be presented in Sec.~\ref{sec_sub_challenges}. The experimental results for maneuver load alleviation, gust load alleviation, as well as simultaneous gust and maneuver load alleviation will be shown in Sec.~\ref{sec_sub_MLA}-\ref{sec_sub_GMLA}. 
\subsection{Experiment Setup}
\label{sec_sub_exp}

\begin{figure}[!h]
\centering
\includegraphics[width=0.88\textwidth]{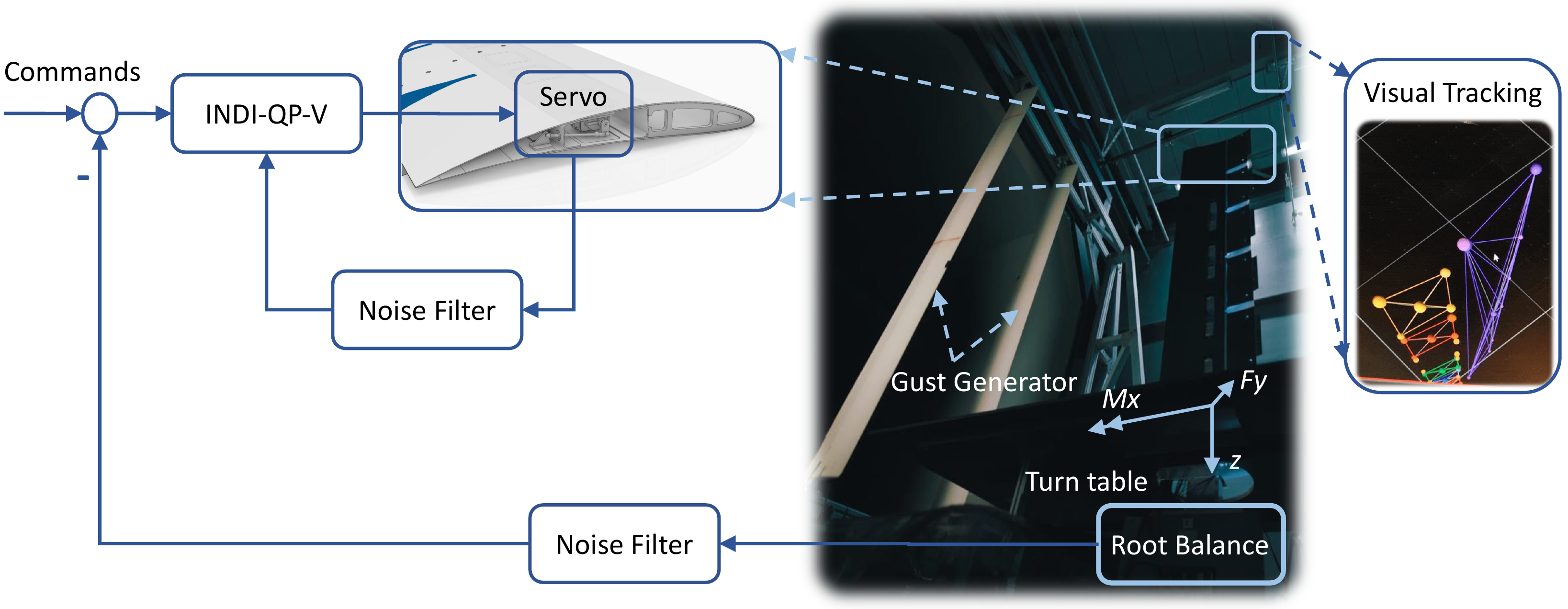}
\caption{A block diagram for experiment setup.}
\label{SmartX_block}
\end{figure}

The experiments were conducted in the Open Jet Facility (OJF) wind tunnel of Delft University of Technology. A two-vane gust generator is installed to produce aerodynamic disturbances at various magnitude and frequencies. The SmartX-Alpha wing has twelve independent servos (Fig.~\ref{fig_morphing_mechanism}), thus $m=12$. In order to alleviate the excessive loads (no matter caused by gusts or maneuvers) without degrading the rigid-body command tracking performance, the load alleviation problems are converted to load reference tracking problems. As discussed in Sec.~\ref{sec_control_algorithms}, the $\boldsymbol y$ in Eq.~(\ref{non_sys}) can be a function of any subset of the physical measured outputs. For load alleviation purposes, choose $\boldsymbol y = [\int F_y, \int M_x]^\mathsf{T}$, where $F_y$ and $M_x$ are the measured wing root shear force and bending moment, respectively (Fig.~\ref{SmartX_block}). Referring to the Theodorsen's theory, given a control surface deflection (a camber morphing for SmartX-Alpha), a half of the circulatory lift gradually build-up, while the rest happens immediately. Therefore, a change in wing camber has direct influences on loads. Accordingly, for the selected inputs and outputs, the vector relative degree is $\boldsymbol{\rho} = [1,1]^\mathsf{T}$. 

Recall Sec.~\ref{sec_control_algorithms}, the only model information needed by INDI is the estimated control effectiveness matrix $\bar{ {\boldsymbol{\mathcal{B}}}}(\boldsymbol x_0)$. For the selected input and output vectors, $\bar{ {\boldsymbol{\mathcal{B}}}}(\boldsymbol x_0) \in \mathbb{R}^{2\times 12}$. In theory, $\bar{ {\boldsymbol{\mathcal{B}}}}$ is a function of states. Nonetheless, as has been proved in Theorems~\ref{theorem_3} and~\ref{theorem_4}, the INDI control can passively resist a wide range of model uncertainties in $\bar{ {\boldsymbol{\mathcal{B}}}}$. Therefore, in the experiment, a constant $\bar{ {\boldsymbol{\mathcal{B}}}}$ matrix identified in the trimmed condition was consistently used by the controller. In this way, the control implementation process was simplified; the robustness of the controller was also tested. 

In~Eq.~(\ref{u_indi_track}), the gain matrix is chosen as $\boldsymbol{K} = \text{diag}\{0.1,0.1\}$. The position constraints for the servos are $\boldsymbol{ u}_{\text{max}} = \boldsymbol{I}_{12\times 1} \cdot 30$~deg, $\boldsymbol{ u}_{\text{min}} = -\boldsymbol{I}_{12\times 1} \cdot 30$~deg. The rate constraints for the servos are $\underline{\boldsymbol{ u}}_{\text{rate}} = - \boldsymbol{I}_{12\times 1} \cdot80$~deg/s, $\bar{\boldsymbol{ u}}_{\text{rate}} = \boldsymbol{I}_{12\times 1} \cdot80$~deg/s. The relative position constrain vector $\bar{\boldsymbol{ u}}_{\text{adj}}\in \mathbb{R}^{11\times 1}$. For $i=1,2,...,11$, when $i$ is an odd number, $\bar{ u}_{\text{adj},i} = 55$~deg; otherwise, $\bar{ u}_{\text{adj},i} = 10$~deg. In Eq.~\eqref{QP_J}, $\sigma$ is chosen as $0.001$ to prioritize $\mathcal{J}_1$. The weighting matrices are chosen as $\boldsymbol{W}_1 = \boldsymbol{I}_{2\times 2}$ and $\boldsymbol{W}_2 = \boldsymbol{I}_{12\times 12}$. A block diagram for the experiment setup is presented in Fig.~\ref{SmartX_block}. 

As shown in Fig.~\ref{SmartX_block}, the SmartX-Alpha wing was vertically mounted on a turn table. The operational point was selected as $V = 15$~m/s, $\alpha =-2.89 $~deg (turn table angle equals 1.00~deg). The three-axes root reaction forces and moments were measured by the OJF External Balance in 1000~Hz. The core component of this balance is a set of strain gauges. For real-world aircraft, strain gauges can also be attached to wing-root structures to provide root reaction forces for feedback control. All the twelve servos were connected to a RS-485 device, communicating serially over the physical USB bus updating at 66.7~Hz. The communication delay was approximately 15~ms. The wing displacements were captured by a visual tracking system (OptiTrack). The local wing loads were measured by embedded strain gauges.  

\subsection{Practical Issues}
\label{sec_sub_challenges}

\subsubsection{Nonlinear Backlash}
\label{subsub_backlash}

\begin{figure}[!h]
\centering
\includegraphics[width=.4\textwidth]{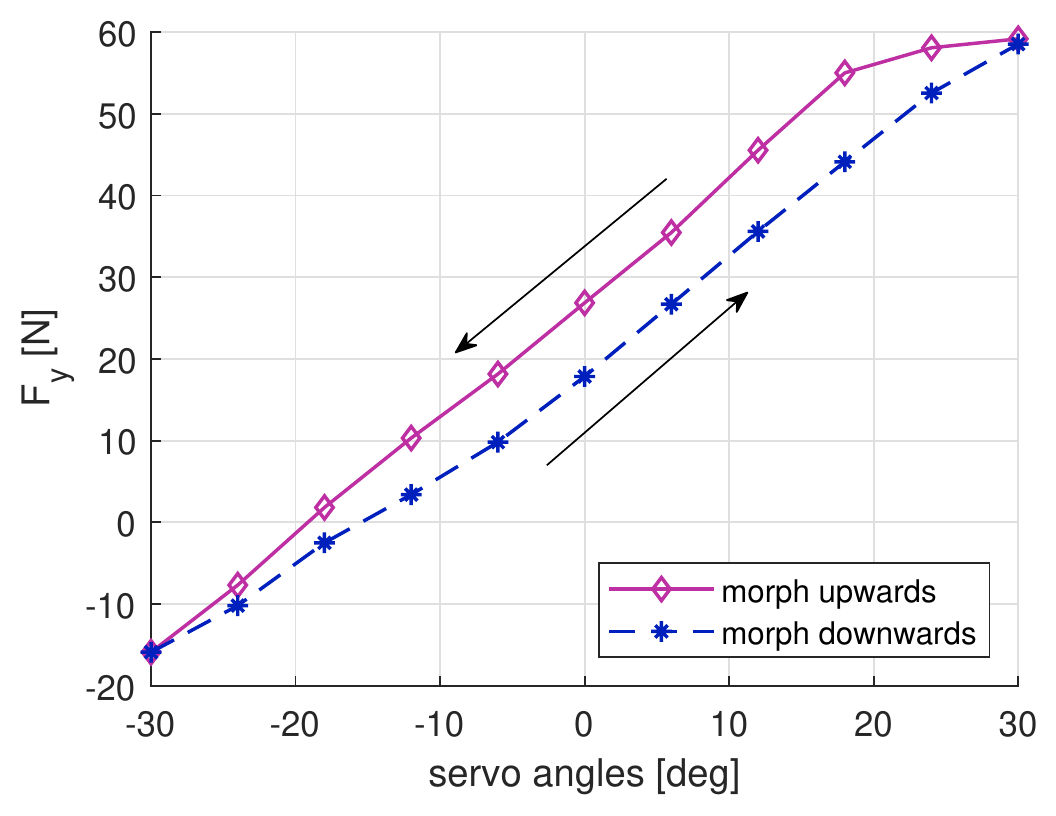}
\caption{Backlash-induced wing root shear force hysteresis loop in the experiment.}
\label{freeplay}
\end{figure}

Backlash is a clearance or lost motion phenomenon in mechanical systems caused by gaps between the mechanical components. Consider a general mechanical linkage; denote the generalized displacement of the driving and driven part as $u$ and $\tau$, respectively. The widely adopted free-play model is~\cite{Gold2008}: $\text{if}~~ u < u_{f_-}, \tau = k_1 (u - u_{f_-}); \text{if}~~ u > u_{f_+}, \tau = k_2 (u - u_{f_+}); \text{otherwise},  \tau = 0$. $k_1>0,k_2>0$ are the linear slopes; $u_{f_+}>0$ and $u_{f_-}<0$ represent the free-play deadband. Actuator free-play can lead to limit cycle oscillations~\cite{Frampton2000}. The backlash nonlinearity is even more challenging~\cite{Campos2000}:
\begin{spacing}{1.3}
\begin{eqnarray}
     \dot\tau = f(\tau,u,\dot u)= \left\{
\begin{array}{ll}
k_1 \dot u,   & \text{if}~~ \dot u < 0~\text{and}~\tau =  k_1 (u - u_{f_-})\\
k_2 \dot u,   & \text{if}~~ \dot u > 0~\text{and}~\tau =  k_2 (u - u_{f_+})\\
0,     & \text{otherwise}
\end{array} \right.
\label{backlash}
\end{eqnarray}
\end{spacing}
Equation~\eqref{backlash} presents a velocity-driven dynamic system. Different from the free-play, $\tau$ in Eq.~\eqref{backlash} is also dependent on the history of $u$. This hysteresis effect was also observed during the experiment. In Fig.~\ref{freeplay}, all the twelve servos execute the same command: starts from 30~deg and gradually reduces to -30~deg (surfaces morph upwards), and then gradually increases to 30~deg (surfaces morph downwards). Figure~\ref{freeplay} shows that due to backlash, the same servo angle settings lead to different force responses in upstroke and downstroke. 

The SmartX-Alpha is the first prototype featuring the distributed morphing TRIC concept. The manufacturing and integration process involved largely handcrafted structural components and manual laminate layup, which inevitably lead to manufacturing imperfections. One of such imperfection was the exact tolerance between the skin and the sliding interface. This gap was filled with a spacer that added additional frictions. Combined with slack in the actuator mechanism and a relatively large stiffness drop between the rigid aluminum pick-up and its attachment to the flexible skin, the pick-up point exhibited local out-of-plane rotations. Moreover, the bottom skin exhibited local bending motions. Consequently, whenever the servo command changes directions, the pick-up point needs to rotate and the bottom skin needs to bend a bit, before the ideal translational sliding actually happens. These phenomena were only discovered during the tightly-scheduled experiment, and were not foreseen by the control designs. Therefore, it was decided to test the robustness of the controller to backlash and friction in this experiment.

\subsubsection{Actuator Dynamics and Fault}
\label{sec_sub_act}
The servos of the SmartX-Alpha are the Volz DA 22-12-4112~\cite{Tigran_smasis}. To identify the servo dynamics, a sweep signal with magnitude of $\pm30$~deg was given to the servo. By analyzing the input and output signals, is was identified that the second-order system $H(s) = \frac{\omega^2}{s^2 + 2 \zeta \omega s + \omega^2}$ can represent the servo dynamics. The identified parameters are $\zeta = 0.71$, $\omega = 16.52~\text{rad/s}$. Consequently, the cut-off frequency of the servo equals 16.35~rad/s (2.60~Hz).

After conducting the control effectiveness identification and before implementing the controllers, the 9$^{\text{th}}$ actuator was non-operational. This failure resulted from adhesive bond failure between the aluminum pick-up and the composite morphing skin. Consequently, the control effectiveness of the 9$^{\text{th}}$ actuator becomes zero. Moreover, since shear forces can still propagate within module four via the composite shell, and propagate to the adjacent module via the elastomer, the control effectiveness of the 8$^{\text{th}}$ and 10$^{\text{th}}$ actuators were also affected. The repair would require to unmount the wind tunnel setup, extract the morphing trailing edge from the wing structure, and wait for new adhesive layer to cure. Given the time constraints, a choice was made to disable the servo and test the robustness of the controller to actuator failures. Also, the control effectiveness identified in the healthy condition were still used in implementations.  

\subsubsection{Colored Noise}
\label{sec_sub_color_noise}

The signals provided by the root balance contain measurement noise. Experimental results show that the measurement noises of $F_y$ and $M_x$ are colored, and also contain considerable energy in the low-frequency range. To reduce the noise energy, the second-order low-pass filter with transfer function $H(s) = \frac{\omega^2}{s^2 + 2 \zeta \omega s + \omega^2}$ is selected. Choosing the filter parameters is a trade-off: a low cut-off frequency leads to better noise attenuation, but causes larger phase lag in the closed-loop system. After several experimental trials, the parameters of the noise filter were chosen as $\zeta = 0.8$, $\omega = 10~\text{rad/s}$. Consequently, the noise filter cut-off frequency equals 8.67~rad/s (1.38~Hz). 
\subsection{Maneuver Load Alleviation}
\label{sec_sub_MLA}

In this subsection, the maneuver load alleviation performance of the proposed controller will be evaluated experimentally. An aircraft symmetric pull-up maneuver is considered. The control objective is to increase lift while reducing wing root bending moment by spanwise lift redistribution.

\begin{figure}[!h]
\centering
\scalebox{1}{
\begin{subfigure}[t]{0.49\textwidth}
\includegraphics[width=\textwidth]{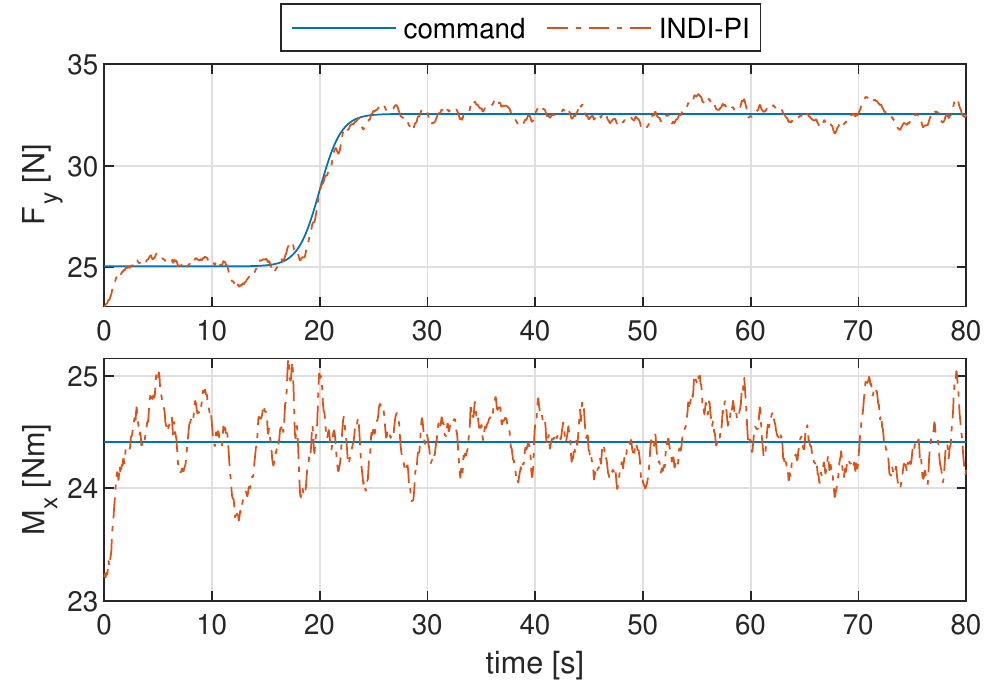}
\caption{Load responses.}
\label{Dec_MLA_PI_loads_Shear30_Bend0}
\end{subfigure}
\begin{subfigure}[t]{0.49\textwidth}
\includegraphics[width=\textwidth]{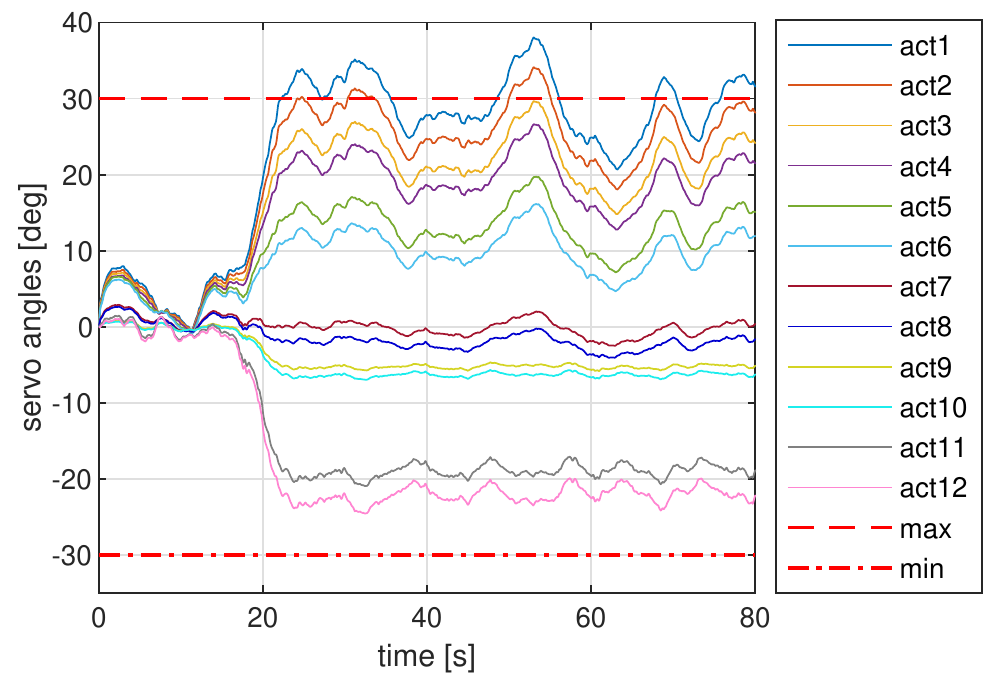}
\caption{Servo angle commands.}
\label{Dec_MLA_PI_servo_Shear30_Bend0}
  \end{subfigure}
}
\caption{Maneuver load alleviation performance of INDI with pseudo inverse control allocation.}
\label{Dec_MLA_PI}
\end{figure}

In Fig.~\ref{Dec_MLA_PI}, $F_y$ is commanded to increase by 30~\%, while $M_x$ is commanded to remain at its trimmed value. A sigmoid function is adopted for a smooth command transition. Figure~\ref{Dec_MLA_PI_loads_Shear30_Bend0} shows that the load commands are tracked in spite of actuator fault, delay, and backlash. Moreover, as illustrated in Fig.~\ref{Dec_MLA_PI_servo_Shear30_Bend0}, the servo at the wing tip (12$^{\text{th}}$) receives negative command (making the wing morph upwards), while the servo command gradually increases from the wing tip to the root. As a consequence, the wing aerodynamic center is moved inboard by the trailing-edge morphing. 

However, neither input constraint nor spanwise servo location is considered in this pseudo inverse control allocation (Sec.~\ref{sec_sub_CA}). Two drawbacks are identified: first, the hardware constraints can be violated (1$^{\text{st}}$ and 2$^{\text{nd}}$ servos in Fig.~\ref{Dec_MLA_PI_servo_Shear30_Bend0}); second, it can cause high tension in the elastomer. For example, at $t= 53.3$~s, the command difference between the 6$^{\text{th}}$ and 7$^{\text{th}}$ servos are 14.2~deg. However, the spanwise distance between theses two servos is only 29.0~mm. This rapid angle change in a short distance can overstretch the elastomer.     

\begin{figure}[!h]
\centering
\scalebox{1}{
\begin{subfigure}[t]{0.49\textwidth}
\includegraphics[width=\textwidth]{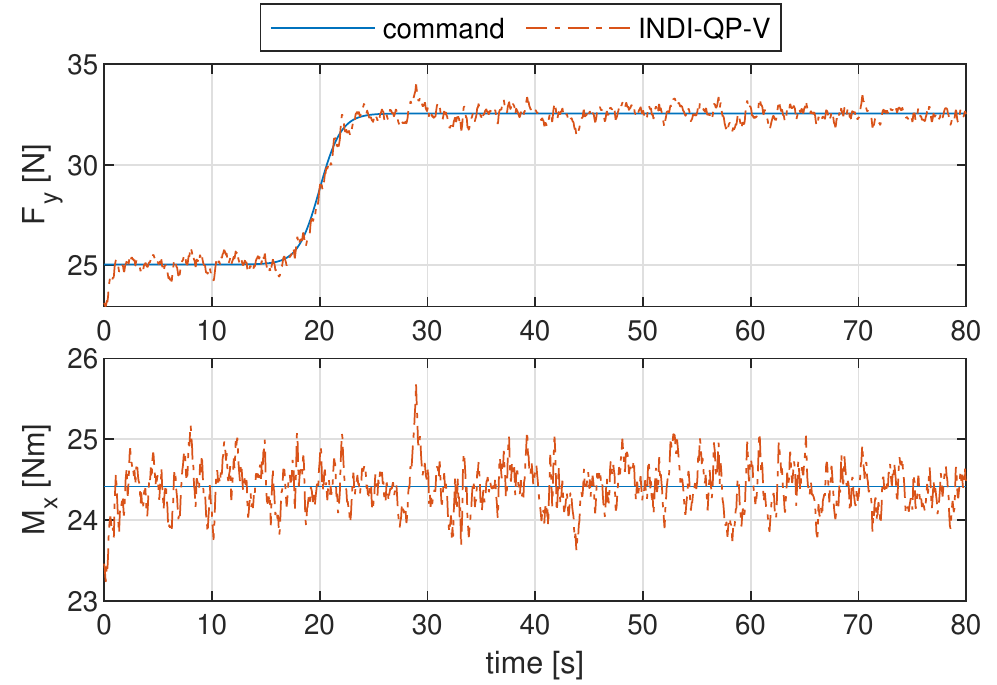}
\caption{Load responses.}
\label{Dec_MLA_QP_V_loads_Shear30_Bend0}
\end{subfigure}
\begin{subfigure}[t]{0.49\textwidth}
\includegraphics[width=\textwidth]{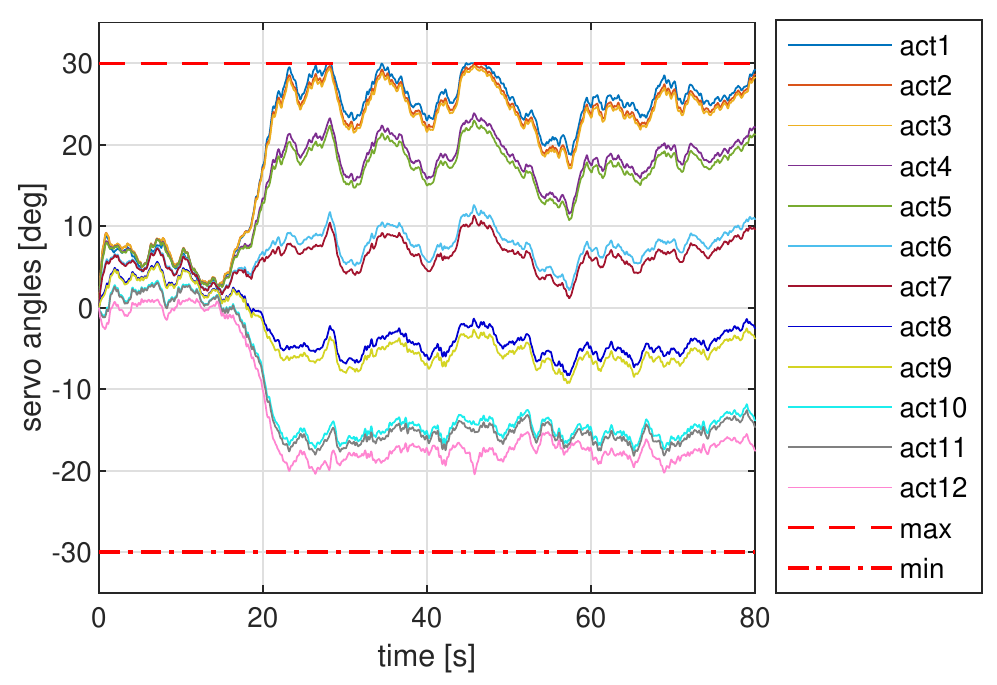}
\caption{Servo angle commands.}
\label{Dec_MLA_QP_V_control_Shear30_Bend0}
  \end{subfigure}
}
\caption{Maneuver load alleviation performance of INDI using quadratic programming and virtual shapes.}
\label{Dec_MLA_QP_V}
\end{figure}

These two drawbacks are overcome in INDI-QP-V, which explicitly considers input constraints and ensures the wing smoothness. Figure~\ref{Dec_MLA_QP_V_loads_Shear30_Bend0} shows that INDI-QP-V increases $F_y$ by 30~\% without amplifying $M_x$. Figure~\ref{Dec_MLA_QP_V_control_Shear30_Bend0} confirms that the input constraints are not violated and inter-modular command gaps are much smaller than the case in Fig.~\ref{Dec_MLA_PI_servo_Shear30_Bend0}.

\begin{figure}[!h]
\centering
\scalebox{1}{
\begin{subfigure}[t]{0.49\textwidth}
\includegraphics[width=\textwidth]{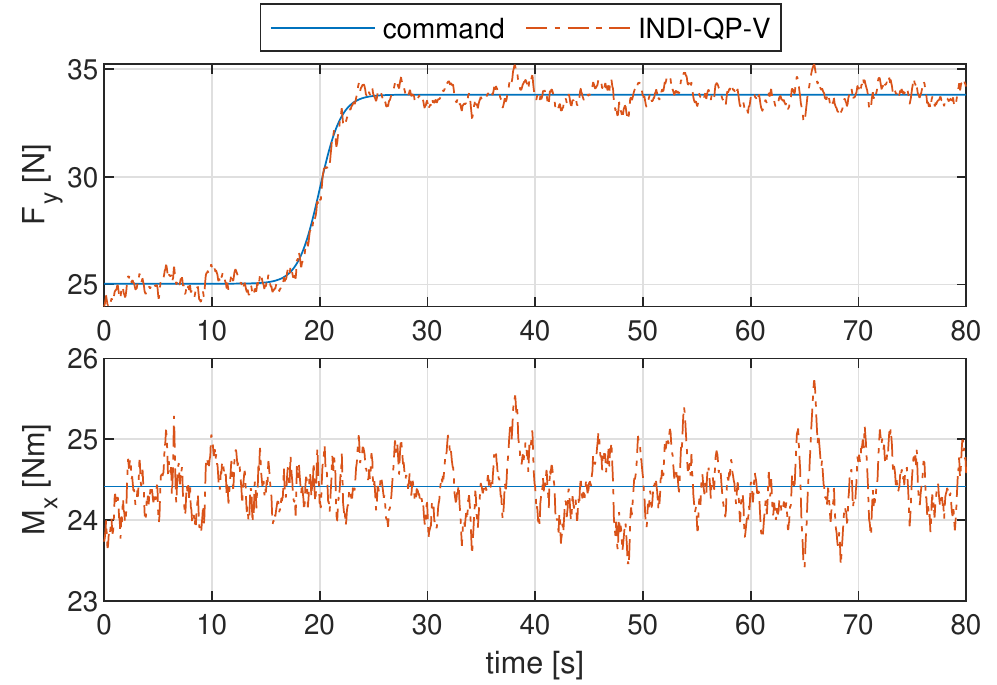}
\caption{Load responses.}
\label{Dec_MLA_QP_V_loads_Shear35_Bend0}
\end{subfigure}
\begin{subfigure}[t]{0.49\textwidth}
\includegraphics[width=\textwidth]{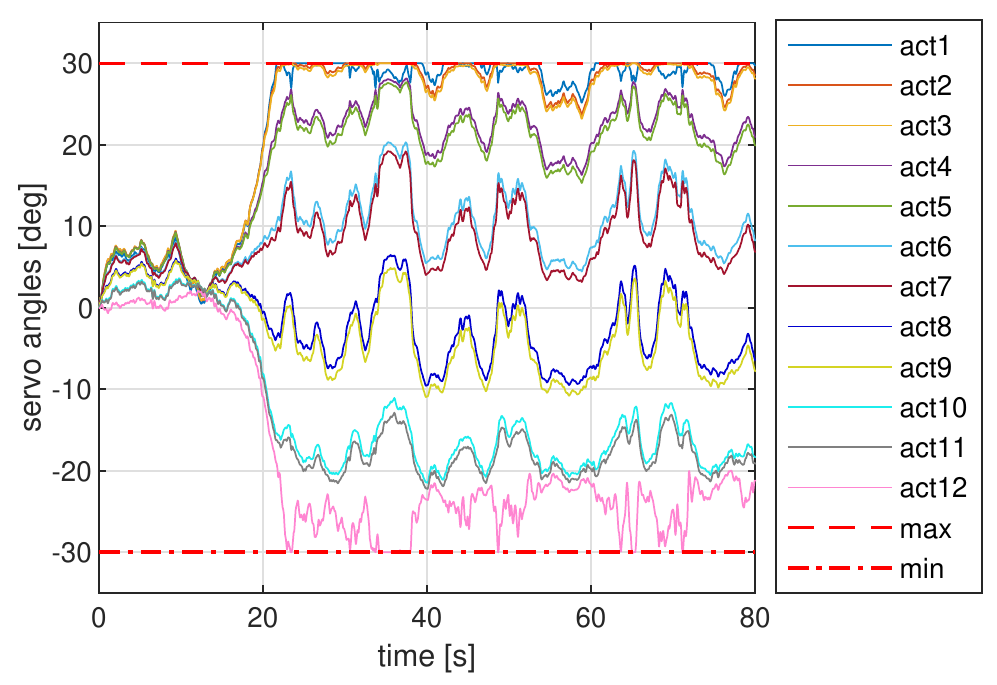}
\caption{Servo angle commands.}
\label{Dec_MLA_QP_V_control_Shear35_Bend0}
  \end{subfigure}
}
\caption{Performance of INDI-QP-V in a challenging maneuver load alleviation task.}
\label{Dec_MLA_QP_V_35}
\end{figure}

To further demonstrate the effectiveness of INDI-QP-V, the controller is asked to increase $F_y$ by 35~\% without raising $M_x$. Since this load alleviation task is more challenging than the previous one, the servo angle commands in Fig.~\ref{Dec_MLA_QP_V_control_Shear35_Bend0} are also saturated more frequently. Nevertheless, the input constraints are not violated; the inter-modular transitions are smooth; the load alleviation mission is also achieved (Fig.~\ref{Dec_MLA_QP_V_loads_Shear35_Bend0}).

\begin{figure}[!h]
\centering
\includegraphics[width=.43\textwidth]{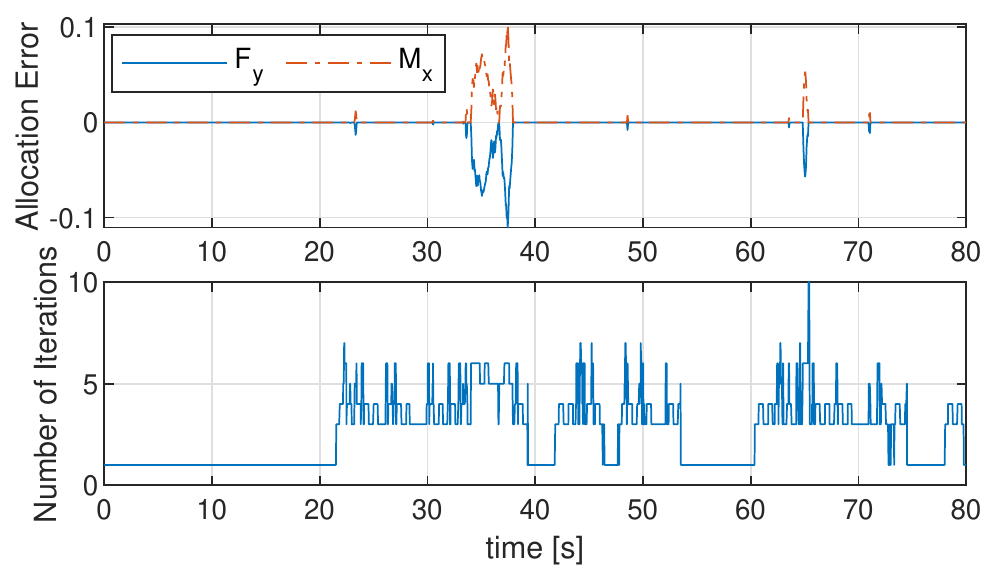}
\caption{Allocation errors and number of iterations of INDI-QP-V in a challenging MLA task.}
\label{Dec_index}
\end{figure}

Figure~\ref{Dec_index} shows that at the majority of time span, $\|\boldsymbol{\varepsilon}_{\text{ca}}\|_2 \leq 1\times 10^{-3}$. When severe saturation occurs, $\|\boldsymbol{\varepsilon}_{\text{ca}}\|_2$ is still bounded by 0.12. Moreover, the allocator converges within one step when there is no saturation, and converges within ten iterations when saturation occurs. In all cases, the computational load is low, and the control commands are realized in real time.

\subsection{Gust Load Alleviation}
\label{sec_seb_GLA}

In Sec.~\ref{sec_sub_MLA}, experimental results have demonstrated that INDI-QP-V is better than INDI-PI. This subsection will present the gust load alleviation effectiveness of INDI-QP-V. To generate each ``1-cos'' gust, the rotational angle of each gust generator vane obeys: $\theta(t) = A_g (1-\cos(2\pi f_g t +\phi))$, where $\phi$ is the phase shift. The corresponding gust angle is $\alpha_g(t) =( A_g/2) (1-\cos(2\pi f_g (t-d_{gw}/V)+\phi ))$, where $d_{gw}$ represents the gust travel distance; $V$ is the nominal wind speed. To test robustness, the gust information was kept unknown to the controller.

Denote the references for $F_y$ and $M_x$ as $F_{y_*}$ and $ M_{x_*}$, respectively. Four performance metrics are introduced: 1) the reduction rate of the maximum value of $F_y - F_{y_*}$; 2) the reduction rate of the root mean square (rms) value of $F_y - F_{y_*}$; 3) the reduction rate of the maximum value of $M_x - M_{x_*}$; 4) the reduction rate of the rms value of $M_x - M_{x_*}$. Take the last performance metric as an example, the reduce rate is calculated as $\frac{(\text{rms}(M_x - M_{x_*}))|_{\text{open}} - (\text{rms}(M_x - M_{x_*}))|_{\text{closed}} }{(\text{rms}(M_x - M_{x_*}))|_{\text{open}}} $, where $(\cdot)|_{\text{open}}$ and $(\cdot)|_{\text{closed}}$ respectively means evaluating $(\cdot)$ in the open-loop or closed-loop condition. 

\begin{figure}[!h]
\centering
\scalebox{1}{
\begin{subfigure}[t]{0.49\textwidth}
\includegraphics[width=\textwidth]{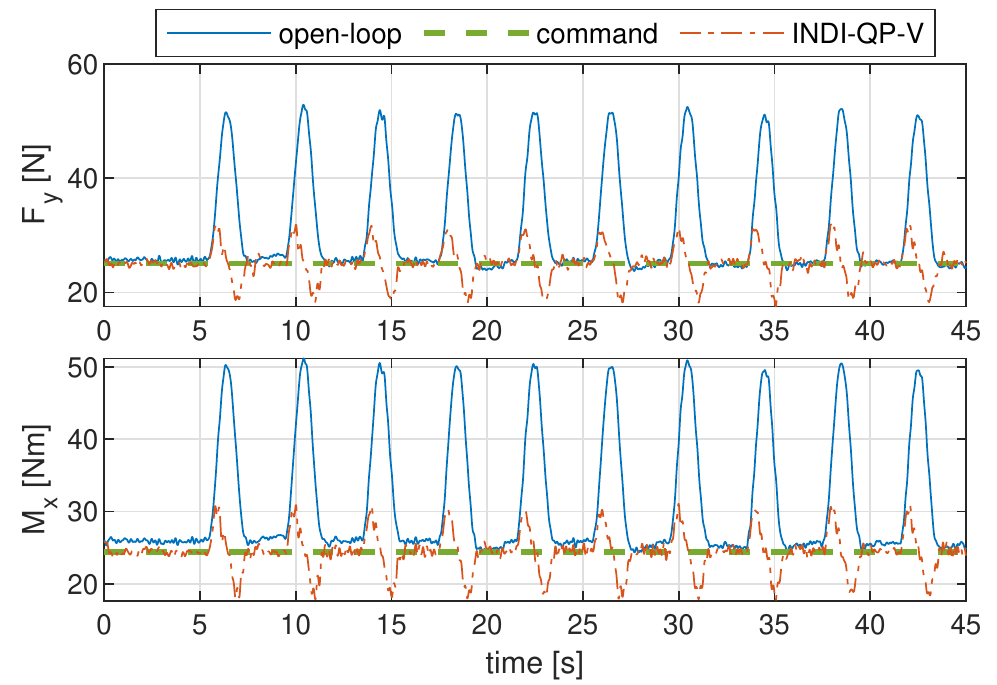}
\caption{Load responses.}
\label{Dec_INDI_GLA_QP_V_05Hz_loads}
\end{subfigure}
\begin{subfigure}[t]{0.49\textwidth}
\includegraphics[width=\textwidth]{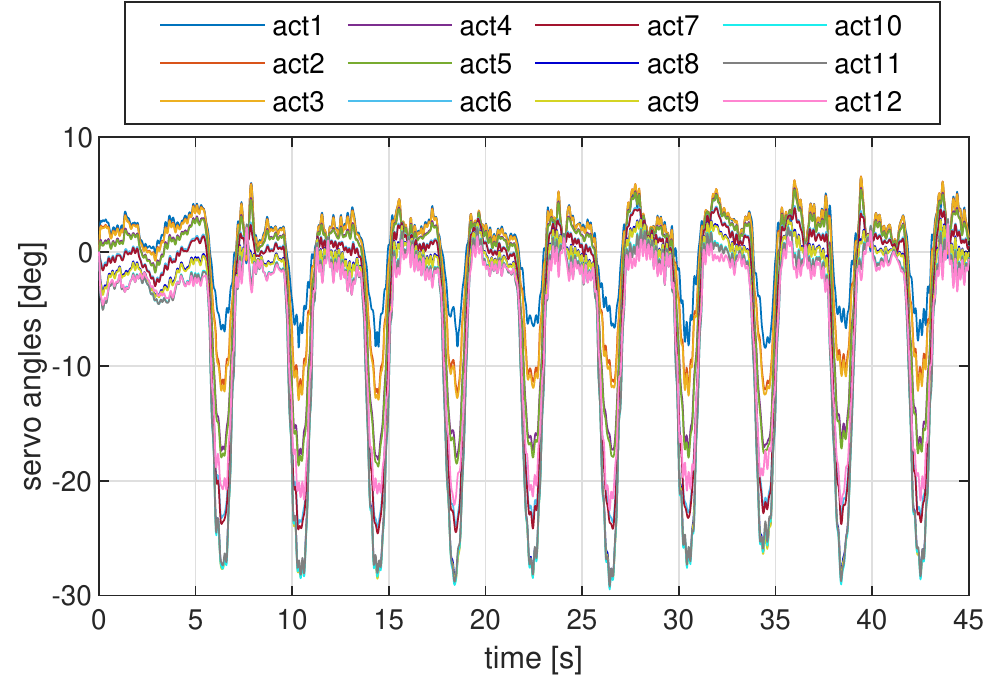}
\caption{Servo angles.}
\label{Dec_INDI_GLA_QP_V_05Hz_servo}
  \end{subfigure}
}
\caption{Load alleviation performance of INDI-QP-V under 0.5~Hz gusts.}
\label{Dec_GLA_QP_V_05Hz}
\end{figure}

In Fig.~\ref{Dec_GLA_QP_V_05Hz}, the gust generator motions obey $A_g = 3.5$~deg and $f_g = 0.5$~Hz. In the open-loop case, the maximum load increments in $F_y$ and $M_x$ are 27.85~N and 26.72~N$\cdot$m, respectively. By using INDI-QP-V, these values are reduced to 6.88~N and 6.64~N$\cdot$m. Over 75~\% of reduction rate is achieved in all the four performance metrics (Table~\ref{table_GLA}). Figure~\ref{Dec_INDI_GLA_QP_V_05Hz_servo} shows that the inter-modular transitions are smooth and no saturation occurs. Because of the colored measurement noises, the measured load variations are non-zero even without gust. When these relatively small variations are fed back to the controller, small oscillatory commands are generated. However, due to backlash (Sec.~\ref{subsub_backlash}), a servo angle change within the deadband has no effect on the morphing surface, which further results in null load change. At the next time step, when the controller ``sees'' the previous command has no effect, a command with higher magnitude will be given to the servo until it moves out of the deadband. These are the physical explanations for the high-frequency oscillations in Fig.~\ref{Dec_INDI_GLA_QP_V_05Hz_servo}.
\begin{table}[!h]
 \begin{center}
  \caption{Gust load reduction rate using INDI-QP-V at various frequencies.}
  \label{table_GLA}
  \begin{tabular}{ccccc}
  \hline\hline
        Frequency [Hz]  & $\max (F_y - F_{y_*})$ & $\text{rms} (F_y - F_{y_*}) $ & $\max (M_x - M_{x_*}) $ & $ \text{rms} (M_x - M_{x_*})$\\ \hline
        0.5  & 75.57~\% & 76.41~\% & 75.16~\% & 77.39~\% \\ 
        1.0  & 56.25~\% & 53.73~\% & 57.04~\% & 56.13~\% \\  
        1.5  & 47.86~\% & 40.80~\% & 47.87~\% & 43.28~\% \\  
        2.0  & 40.52~\% & 29.23~\% & 40.47~\% & 32.58~\% \\ 
        2.5  & 25.25~\% & 19.49~\% & 27.34~\% & 24.23~\% \\ 
        3.0  & 15.26~\% & 14.83~\% & 18.53~\% & 20.77~\% \\ 
        3.5  & 7.52~\%  & 6.12~\%  & 7.14~\%  & 14.29~\% \\ 
        4.0  & 8.83~\%  & -1.44~\%  & 10.28~\% &6.98~\%  \\ 
        4.5  & -0.96~\%  & -6.77~\%  & 5.79~\%  &4.24~\% \\
       \hline
        \hline
  \end{tabular}
 \end{center}
\end{table}

\begin{figure}[!h]
\centering
\scalebox{1}{
\begin{subfigure}[t]{0.49\textwidth}
\includegraphics[width=\textwidth]{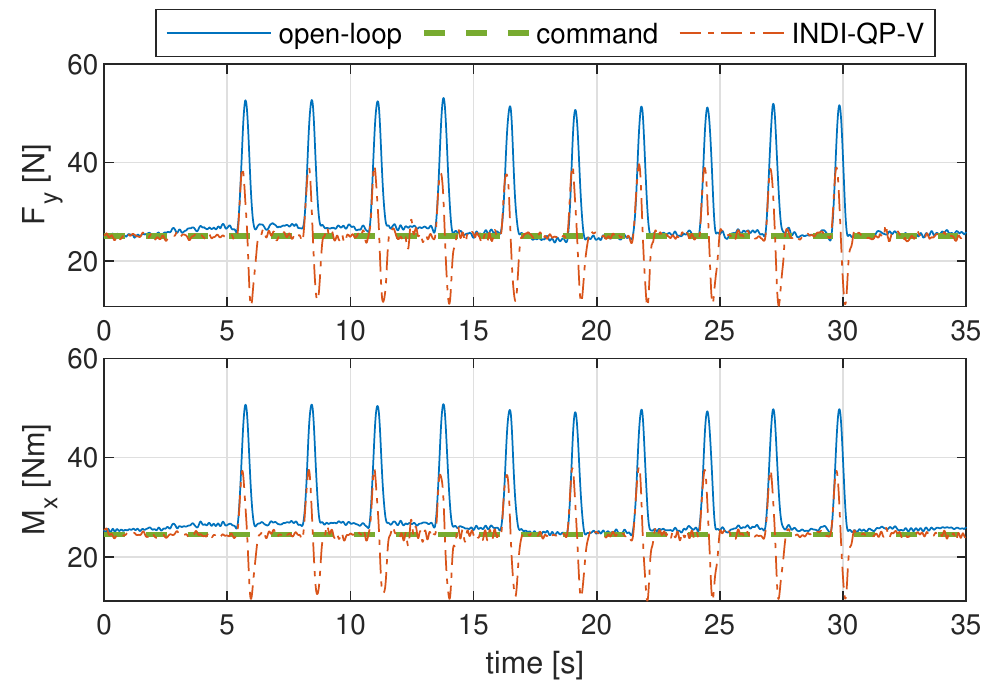}
\caption{Load responses.}
\label{Dec_INDI_GLA_QP_V_15Hz_loads}
\end{subfigure}
\begin{subfigure}[t]{0.49\textwidth}
\includegraphics[width=\textwidth]{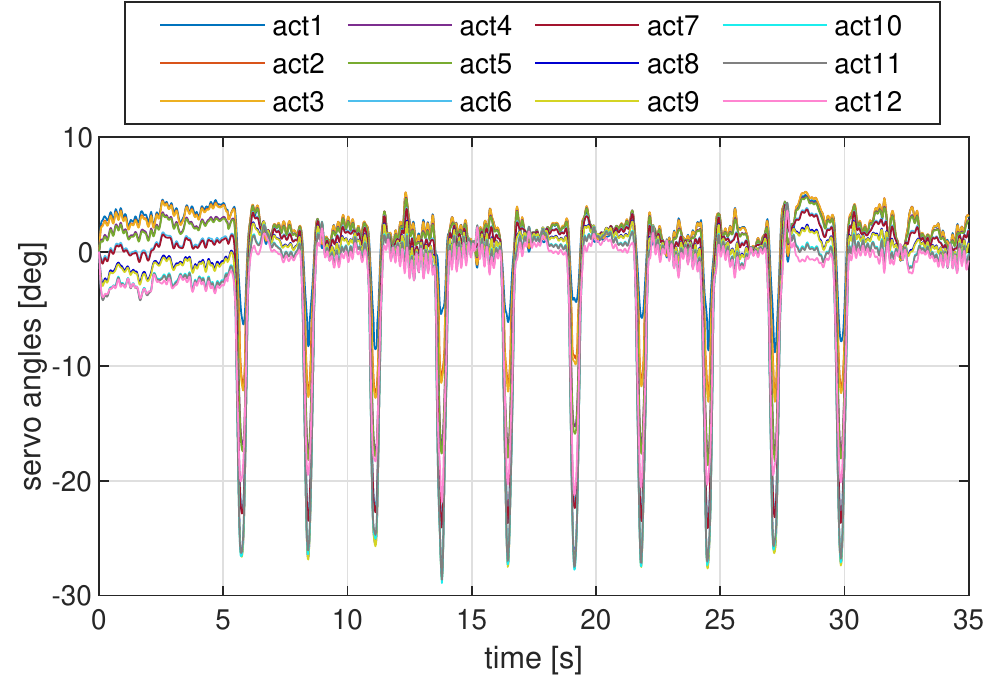}
\caption{Servo angles.}
\label{Dec_INDI_GLA_QP_V_15Hz_servo}
  \end{subfigure}
}
\caption{Load alleviation performance of INDI-QP-V under 1.5~Hz gusts.}
\label{Dec_GLA_QP_V_15Hz}
\end{figure}

Figure~\ref{Dec_GLA_QP_V_15Hz} illustrates the open- and closed-loop load responses when $A_g = 3.5$~deg and $f_g = 1.5$~Hz. In the open-loop case, the maximum load increments in $F_y$ and $M_x$ are 28.09~N and 26.36~N$\cdot$m, respectively. With the help of INDI-QP-V, these values are respectively reduced to 14.65~N and 13.74~N$\cdot$m. Table~\ref{table_GLA} shows that more than 40~\% of load reductions are achieved in all the four performance metrics. Comparing Fig.~\ref{Dec_GLA_QP_V_15Hz} with Fig.~\ref{Dec_GLA_QP_V_05Hz}, we can see that the alleviation performance degrades with the increase in gust frequency. Moreover, the closed-loop load responses in Fig.~\ref{Dec_INDI_GLA_QP_V_15Hz_loads} are more lagged behind than those in Fig.~\ref{Dec_INDI_GLA_QP_V_05Hz_loads}.

Table~\ref{table_GLA} summarizes the load reduction rates of INDI-QP-V. Note in all cases, the control gains and $A_g$ remain consistent. The reduction rates are over 75~\% percent when $f_g = 0.5$~Hz, but reduces to around 20~\% when $f_g = 2.5$~Hz. When $f_g$ further increases to 4.5~Hz, the rms value of $F_y - F_{y_*}$ is even higher in the closed-loop condition. The main reason for this performance degradation is the phase lag in the closed-loop system. Recall Sec.~\ref{sec_sub_challenges}, the cut-off frequencies of the servo and the noise filter are 2.60~Hz and 1.38~Hz, respectively. These lead to large phase lags in the high-frequency range, which further results in the performance deterioration. To improve the performance, we can use less noisy sensors, faster servos, and disclose the gust information to the controller if onboard gust sensing is available.

\subsection{Simultaneous Gust and Maneuver Load Alleviation}
\label{sec_sub_GMLA}

In the literature, GLA and MLA are usually seen as two research topics. However, during real flights, instead of classifying loads by their causes, it is more meaningful to achieve the necessary loads for performing maneuvers while neutralizing the excessive loads (no matter induced by maneuvers or gusts). INDI-QP-V is a good candidate to achieve this goal. As presented in Sec.~\ref{sec_sub_exp}, the real-world load alleviation task is seen as a load command tracking problem by INDI-QP-V. Consequently, by minimizing the error between the commanded and real loads, simultaneous gust and maneuver load alleviation can be realized. This design also ensures the task applicability. In fact, in the experiments of MLA (Sec.~\ref{sec_sub_MLA}), GLA (Sec.~\ref{sec_seb_GLA}), and simultaneous GLA and MLA (Sec.~\ref{sec_sub_GMLA}), only the load commands are task-dependent; there is no need to change the control architecture nor the control parameters.
\begin{figure}[!h]
\centering
\scalebox{1}{
\begin{subfigure}[t]{0.49\textwidth}
\includegraphics[width=\textwidth]{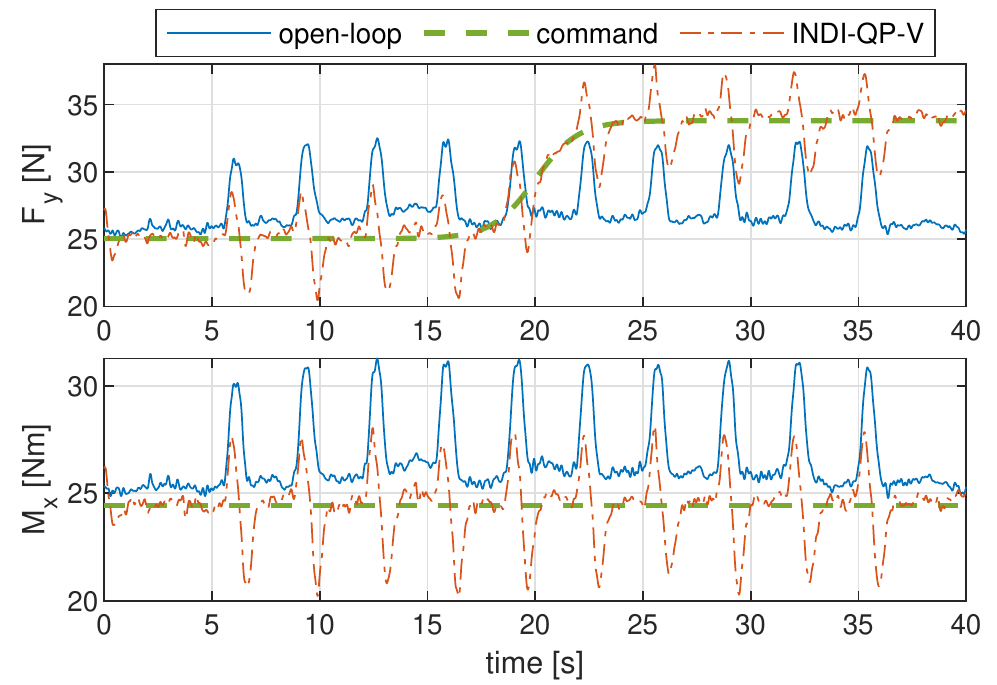}
\caption{Load responses.}
\label{Dec_INDI_MGLA_Shear35_Bend0_loads}
\end{subfigure}
\begin{subfigure}[t]{0.49\textwidth}
\includegraphics[width=\textwidth]{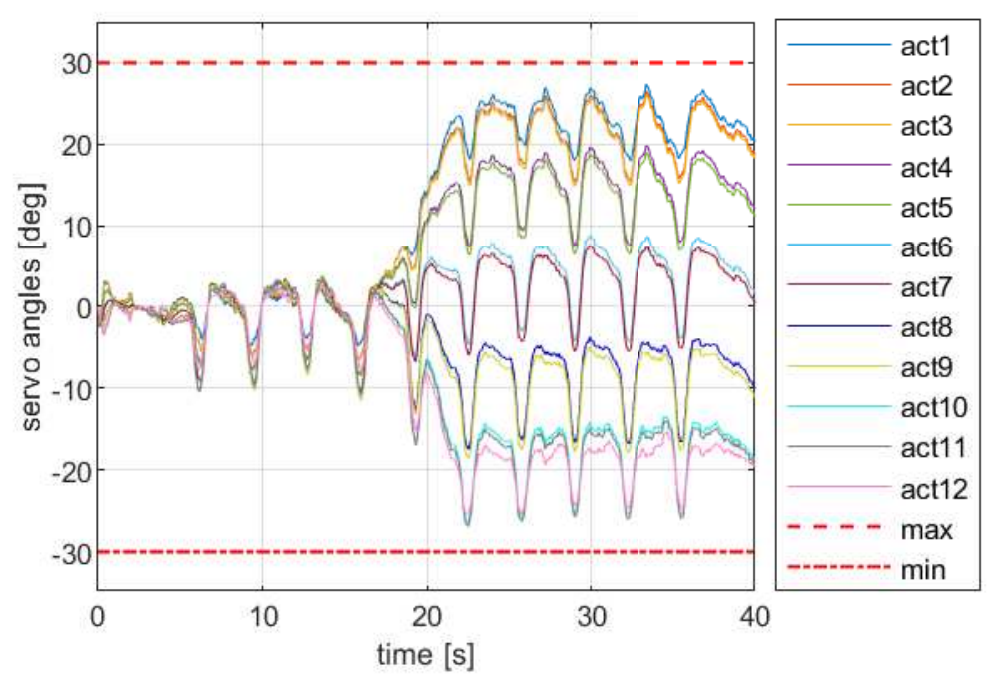}
\caption{Servo angles.}
\label{Dec_INDI_MGLA_Shear35_Bend0_servo}
  \end{subfigure}
}
\caption{Simultaneous maneuver and gust load alleviation performance of INDI-QP-V.}
\label{INDI_MGLA}
\end{figure}

Figure~\ref{INDI_MGLA} presents the experimental results for simultaneous gust and maneuver load alleviation. $F_y$ is commanded to increase by 35~\% for achieving a pull-up maneuver, while $M_x$ is asked to stay at its nominal value in spite of the maneuver and gusts ($A_g = 1$~deg and $f_g = 1$~Hz). Figure~\ref{Dec_INDI_MGLA_Shear35_Bend0_loads} demonstrates that INDI-QP-V is able to alleviate the excessive loads. Using the performance metrics (Sec.~\ref{sec_seb_GLA}), the maximum and rms values of $F_y - F_{y_*}$ are respectively reduced by 44.31~\% and 67.76~\%; the maximum and rms values of $M_x - M_{x_*}$ are reduced by 45.58~\% and 46.35~\%, respectively. After $t=17$~s, the outboard wing starts to morph upwards while the inboard wing begins to morph downwards for spanwise lift redistribution (Fig.~\ref{Dec_INDI_MGLA_Shear35_Bend0_servo}). Moreover, on top of the redistributive motions, the wing actively morphs upwards to reduce the gust-induced loads. Furthermore, the quadratic programming control allocator ensures no saturation occurs; virtual shape functions realize smooth wing shape at every moment. 

\section{Comparisons with Linear Quadratic Gaussian Control}
\label{sec_LQG}

In Sec.~\ref{sec_experiments}, experimental results have demonstrated the effectiveness of INDI-QP-V in GLA, MLA and simultaneous GLA and MLA tasks. In the literature, the Linear Quadratic Gaussian (LQG) control is one of the most popular methods for load alleviation~\cite{Vartio2008a,Nguyen2017,Ferrier2018,Blelloch1990RobustLQ}. Therefore, the proposed INDI-QP-V control will be compared to LQG control. 


\subsection{LQG Control Design}
\label{sec_sub_LQG_design}
The LQG control is essentially a combination of a Kalman filter for state estimation and a linear-quadratic regulator (LQR) for stabilization. The linearized SmartX-Alpha dynamics are: $\dot {\boldsymbol{x}} = \boldsymbol{A} \boldsymbol{x} + \boldsymbol{B} \boldsymbol{u} + \boldsymbol{B}_g \alpha_g,~
\boldsymbol{y} = \boldsymbol{C} \boldsymbol{x} + \boldsymbol{D} \boldsymbol{u} $, where $\alpha_g$ is the gust input angle. $\boldsymbol{y}  = [ \boldsymbol{y}_b^\mathsf{T}, \boldsymbol{y}_a^\mathsf{T}]^\mathsf{T}$, with $\boldsymbol{y}_b$ includes $F_y$ and $M_x$, and $\boldsymbol{y}_a$ denotes the wing accelerations measurements.  

First, assume the states are known, design an LQR to make $\boldsymbol{y}_b$ track its reference $\boldsymbol{y}_{r}$. The LQR design requires the estimated system model: $ \bar{\boldsymbol{A}}, \bar{\boldsymbol{B}}, \bar{\boldsymbol{C}}, \bar{\boldsymbol{D}} $. In view of the benefits of using virtual shape functions (Sec.~\ref{sec_sub_VC}), the following transformation is also adopted by LQR: $\boldsymbol{u} = \boldsymbol {\Phi}_{\bar{x}_s}\boldsymbol{u}_v$. Design an LQR for the following augmented system:
\begin{spacing}{1.3}
\begin{equation}
  \begin{bmatrix}
  \dot {\boldsymbol{x}} \\ 
\boldsymbol y_b -\boldsymbol{y}_r
  \end{bmatrix}  
  = \begin{bmatrix}
 \bar {\boldsymbol{A}}  & \boldsymbol{0} \\
\bar{\boldsymbol{C}}_b & \boldsymbol{0}
  \end{bmatrix}
  \begin{bmatrix}
  {\boldsymbol{x}} \\  
  \int ( \boldsymbol y_b -\boldsymbol{y}_r )
  \end{bmatrix}+
  \begin{bmatrix}
  \bar{\boldsymbol{B}} \\
  \bar {\boldsymbol{D}}_b
  \end{bmatrix} \boldsymbol {\Phi}_{\bar{x}_s}\boldsymbol{u}_v  +
  \begin{bmatrix}
  \boldsymbol{0} \\
  - \boldsymbol {y}_{r}
  \end{bmatrix}
  \label{augmented_LQR}
\end{equation}
\end{spacing}
\noindent where $\bar{\boldsymbol{C}}_b$ and $\bar{\boldsymbol{D}}_b$ respectively equals the first two rows of $\bar{\boldsymbol{C}}$ and $\bar{\boldsymbol{D}}$ (corresponding to $\boldsymbol{y}_b$). Define the augmented state vector as $\boldsymbol{X} = [\boldsymbol{x}^\mathsf{T}, \int ( \boldsymbol y_b -\boldsymbol{y}_r )^\mathsf{T}]^\mathsf{T}$. Denote Eq.~\eqref{augmented_LQR} as $\dot{\boldsymbol{X}} = \boldsymbol A_{\text{aug}} \boldsymbol{X} +\boldsymbol B_{\text{aug}} \boldsymbol{u}_v + \boldsymbol{y}_{r,\text{aug}} $. The cost function is $\mathcal{J}_6 = \lim \frac{1}{2}\int_0^{\infty} [ \boldsymbol{X}^\mathsf{T} \boldsymbol{ Q} \boldsymbol{X} + \boldsymbol{u}_v^\mathsf{T} \boldsymbol{R} \boldsymbol{u}_v]~ \text{d}t$. The resulting optimal control input is $\boldsymbol{u}_v = \boldsymbol{K}_X \boldsymbol{X} + \boldsymbol{K}_r \boldsymbol{y}_{r,\text{aug}} $, $ \boldsymbol{K}_X =-\boldsymbol{R}^{-1} \boldsymbol{B}^\mathsf{T}_{\text{aug}} \boldsymbol{S} $, $ \boldsymbol{K}_r =-  \boldsymbol{R}^{-1} \boldsymbol{B}_{\text{aug}}^\mathsf{T}(\boldsymbol{S}\boldsymbol{B}_{\text{aug}} \boldsymbol{R}^{-1} \boldsymbol{B}_{\text{aug}} ^\mathsf{T}- \boldsymbol{A}_{\text{aug}}^\mathsf{T} )^{-1}\boldsymbol{S} $, in which $\boldsymbol{S}$ is the solution of the associated Riccati equation.

Second, design a Kalman filter for $\dot {\boldsymbol{x}} = \bar{\boldsymbol{A}} \boldsymbol{x} + \bar{\boldsymbol{B}} \boldsymbol{u} + \bar{\boldsymbol{G}} \boldsymbol w,~
\boldsymbol{y} = \bar{\boldsymbol{C}} \boldsymbol{x} + \bar{\boldsymbol{D}} \boldsymbol{u}  +  \bar{\boldsymbol{H}}\boldsymbol w + \boldsymbol v$. The process noise $\boldsymbol w$ and measurement noise $\boldsymbol v$ are assumed to be white. They also satisfies $E(\boldsymbol w \boldsymbol w^\mathsf{T}) = \boldsymbol{Q}_k$, $E(\boldsymbol v \boldsymbol v^\mathsf{T}) = \boldsymbol{R}_k$, $E(\boldsymbol w \boldsymbol v^\mathsf{T}) = \boldsymbol{N}_k$. Design a dynamic system $\dot {\hat{\boldsymbol{x}}} = \bar{\boldsymbol{A}} \hat{\boldsymbol{x}} + \bar{\boldsymbol{B}} \boldsymbol{u} + \boldsymbol{L} \left(\boldsymbol{y} - \bar{\boldsymbol{C}} \hat{\boldsymbol{x}} - \bar{\boldsymbol{D}} \boldsymbol{u}   \right)$, where $\boldsymbol{L}  $ is the optimal Kalman gain, then $\hat{\boldsymbol{x}}\rightarrow {\boldsymbol{x}}$ as $ t \rightarrow \infty$. 

Finally, integrate the LQR controller with the Kalman filter state observer, the resulting LQG control input is $\boldsymbol{u} =\boldsymbol {\Phi}_{\bar{x}_s}\boldsymbol{u}_v = \boldsymbol {\Phi}_{\bar{x}_s} \boldsymbol{K}_X [\hat{\boldsymbol{x}}^\mathsf{T}, \int ( \bar{\boldsymbol{C}}_b \hat{\boldsymbol{x}} + \bar{\boldsymbol{D}}_b \boldsymbol{u} -\boldsymbol{y}_r )^\mathsf{T}]^\mathsf{T} + \boldsymbol {\Phi}_{\bar{x}_s} \boldsymbol{K}_r \boldsymbol{y}_{r,\text{aug}} $.

\subsection{Theoretical Comparisons}

The first difference between INDI-QP-V and LQG is that INDI-QP-V is a nonlinear control method. By contrast, even though LQG can be applied to nonlinear systems, the closed-loop stability is only guaranteed locally. Extending LQG to a wider state definition domain requires the gain-scheduling method. However, the gain-scheduled LQG is tedious to tune; its stability also heavily depends on the linearization density, and cannot be ensured in general cases. 

Second, the robustness of LQG to model uncertainties and external disturbances is not guaranteed. Additional methods such as the loop transfer recovery (LTR) are required to enhance its robust stability~\cite{green2012linear}. Besides, the white noise assumptions can hardly be met in reality. When the uncertainties and disturbances have low-frequency component, although the Kalman filter can be applied, its estimation accuracy degrades. In contrast to LQG, the sensor-based incremental control itself already has shown strong robustness against uncertainties and disturbances~\cite{Sun2020,Wang2019b}.

Last but not least, it is easier to implement INDI-QP-V. The only model information needed by INDI-QP-V is the control effectiveness matrix $\bar{ {\boldsymbol{\mathcal{B}}}}$, while LQG requires an estimation of the complete system dynamic matrices $ \bar{\boldsymbol{A}}, \bar{\boldsymbol{B}}, \bar{\boldsymbol{C}}, \bar{\boldsymbol{D}} $. The tuning of INDI-QP-V is also easier than LQG. All the gains required by INDI-QP-V have physical meanings. On the contrary, when uncertainties and disturbances present, the tuning of $ \boldsymbol{Q}_k, \boldsymbol{R}_k,  \boldsymbol{N}_k$ in LQG is not straightforward. 

As a matter of fact, the tedious system identification (for $ \bar{\boldsymbol{A}}, \bar{\boldsymbol{B}}, \bar{\boldsymbol{C}}, \bar{\boldsymbol{D}} $) and tuning processes were the main barriers in LQG implementation. When an LQR controller was designed based on the identified model and then integrated with a Kalman filter, the resulting LQG performed poorly in our experiment. The identification and tuning took much longer time than planned, and eventually the implemented LQG was not successful within the time limit. On the contrary, we only spent one morning for identifying the $\bar{{\boldsymbol{\mathcal{B}}}}$ needed by INDI-QP-V. The gain tuning of INDI-QP-V was also straightforward. The entire hardware implementation of INDI-QP-V on the SmartX-Alpha was achieved within a day. 

\subsection{Load Alleviation Performance Comparisons}

Since the LQG control did not work in the experiment within the time limit owing to its tedious model identification and tuning process. There is no valid experimental data for LQG. In this subsection, we would like to compare the performance of LQG and INDI-QP-V in the simulation environment. The simulation model was identified from the experimental data. The control parameters of INDI-QP-V are kept the same with those used in the experiment. 

\begin{figure}[!h]
\centering
\scalebox{1}{
\begin{subfigure}[t]{0.49\textwidth}
\includegraphics[width=\textwidth]{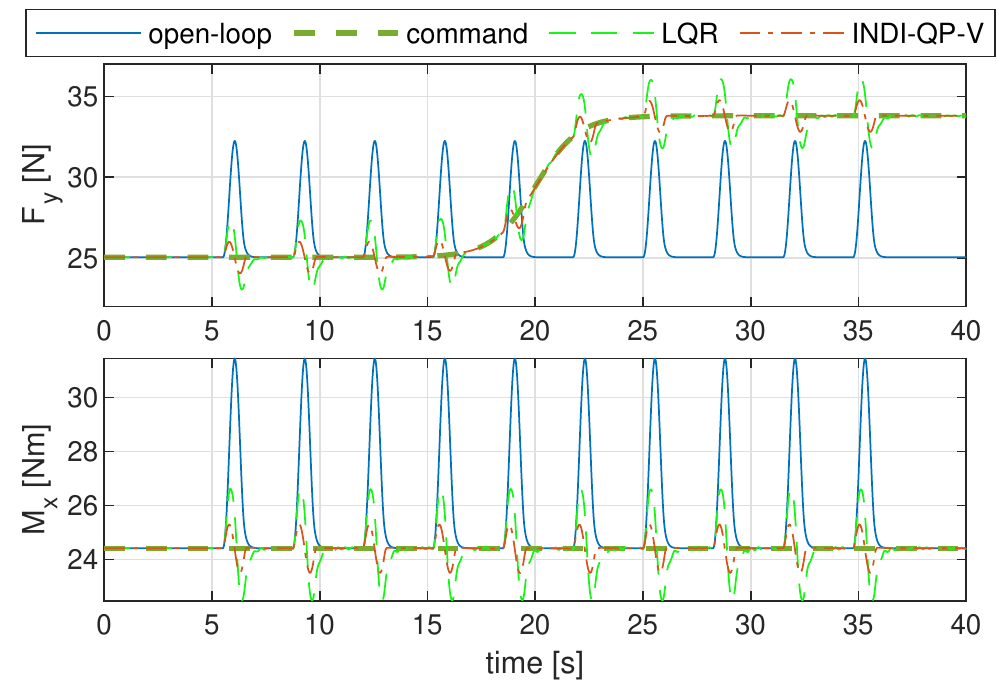}
\caption{Load responses.}
\label{Dec_sim_INDI_LQR_GLMA_loads}
\end{subfigure}
\begin{subfigure}[t]{0.49\textwidth}
\includegraphics[width=\textwidth]{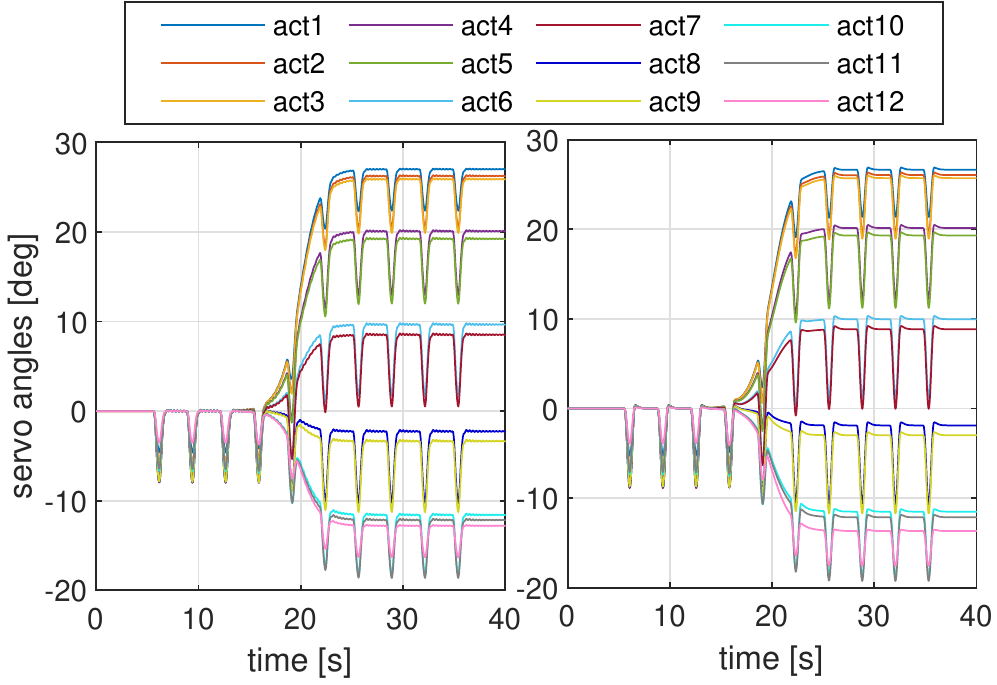}
\caption{Servo angles: LQR (left) and INDI-QP-V (right).}
\label{Dec_sim_INDI_LQR_GLMA_servo}
  \end{subfigure}
}
\caption{Simultaneous gust and maneuver load alleviation without measurement noise.}
\label{Dec_sim_no_noise}
\end{figure}

The comparisons start with an ideal case, where measurement noise, actuator fault and backlash are not included yet. More importantly, the state information is assumed to be known (LQG degrades to LQR). $\boldsymbol{Q}$ is designed as a partitioned matrix, with the upper left matrix equals $\boldsymbol{C}_b^\mathsf{T}\boldsymbol{C}_b$, the lower right matrix equals $10\cdot \boldsymbol{I}_{2\times 2}$, and the rest are equal to zero. $\boldsymbol{R} = 260\cdot \boldsymbol{I}_{5\times 5}$. In Fig.~\ref{Dec_sim_no_noise} the aircraft is asked to perform a pull-up maneuver in a gust field ($A_g = 1$~deg and $f_g = 1$~Hz). Although both controllers can make the wing follow the load commands, INDI-QP-V has better load alleviation performance. The performance metrics are summarized in the second row of Table~\ref{table_GMLA_LQG_INDI} (Sim no noise (Fig.~\ref{Dec_sim_no_noise})). It can be seen that the reduction rate of LQG is above 63~\% while INDI-QP-V reduces loads by more than 85~\%. 
\begin{table}[!h]
\newcommand{\tabincell}[2]{\begin{tabular}{@{}#1@{}}#2\end{tabular}}
 \begin{center}
  \caption{Simultaneous gust and maneuver load reduction rate of INDI-QP-V and LQG.}
   \label{table_GMLA_LQG_INDI}
   \scalebox{0.88}{
  \begin{tabular}{lcccccccc}
  \hline\hline
        \multirow{2}{*}{Conditions}  & \multicolumn{2}{c}{$\max (F_y - F_{y_*})$ }& \multicolumn{2}{c}{$\text{rms} (F_y - F_{y_*}) $} &\multicolumn{2}{c} {$\max (M_x - M_{x_*}) $} & \multicolumn{2}{c}{$ \text{rms} (M_x - M_{x_*})$}\\ \cline{2-9}
        & INDI-QP-V & LQG & INDI-QP-V & LQG & INDI-QP-V & LQG & INDI-QP-V & LQG \\  \hline
    Experiment (Fig.~\ref{INDI_MGLA}) & 44.31~\% &  -  & 67.76~\% & - & 45.58~\% & - & 46.35~\% & -\\ 
    Sim no noise (Fig.~\ref{Dec_sim_no_noise}) & 86.65~\% &  68.25~\% &  93.95~\% &  85.44~\% &  87.48~\% &  68.60~\% &  85.71~\% & 63.73\% \\ 
    Sim with noise (Fig.~\ref{Dec_sim_with_noise}) &
 39.99~\% &  -11.41~\% &  73.99~\% &  63.19~\% &  50.90~\% &  3.71~\% &  35.95~\% & 4.19~\% \\ 
   \tabincell{l}{Sim with noise, fault,\\ and backlash (Fig.~\ref{Dec_sim_everything})}& 25.99~\% &  -22.91~\% &  68.08~\% &  59.47~\% &  38.74~\% &  -12.56~\% &  19.21~\% & -14.24~\% \\ \hline\hline 
  \end{tabular}}
 \end{center}
\end{table}

\begin{figure}[!h]
\centering
\scalebox{1}{
\begin{subfigure}[t]{0.49\textwidth}
\includegraphics[width=\textwidth]{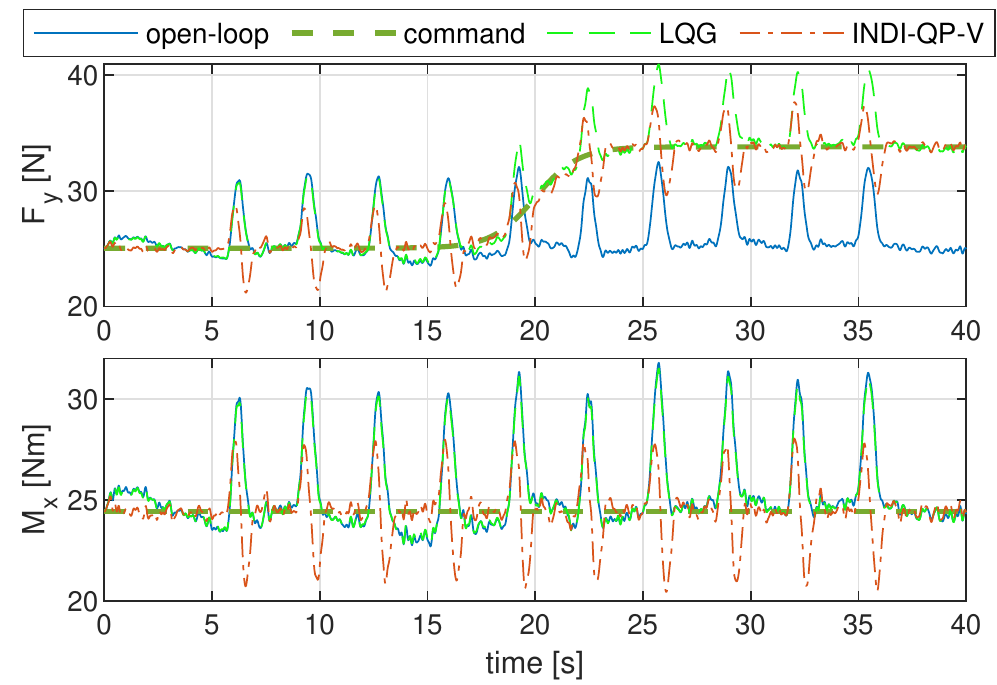}
\caption{Load responses.}
\label{Dec_sim_INDI_LQR_GLMA_with_noise_loads}
\end{subfigure}
\begin{subfigure}[t]{0.49\textwidth}
\includegraphics[width=\textwidth]{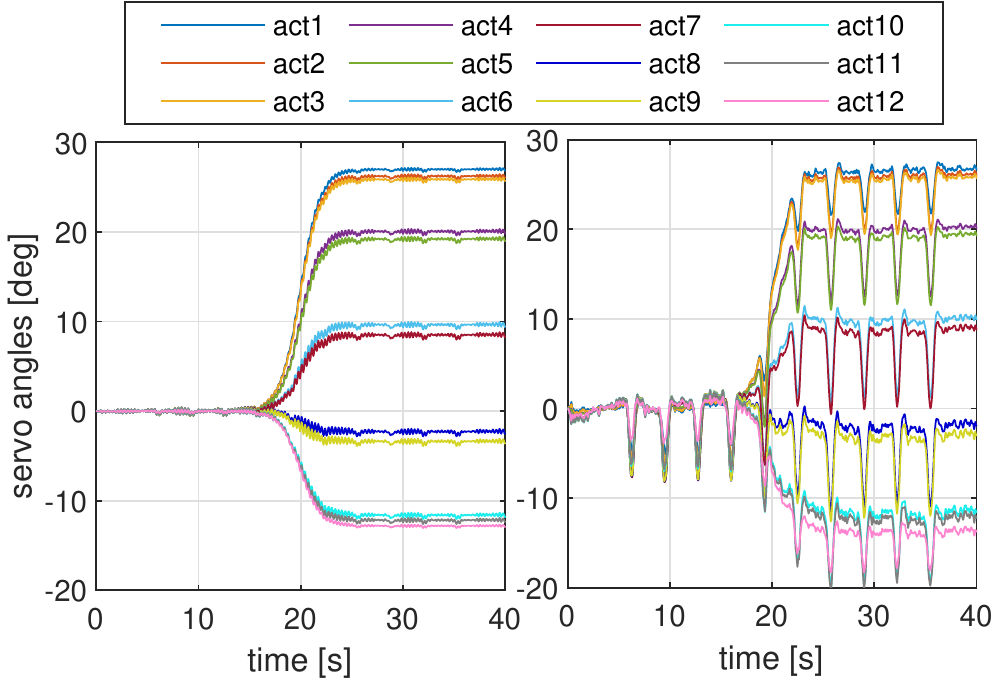}
\caption{Servo angles: LQG (left) and INDI-QP-V (right).}
\label{Dec_sim_INDI_LQR_GLMA_with_noise_servo}
  \end{subfigure}
}
\caption{Simultaneous gust and maneuver load alleviation with measurement noise.}
\label{Dec_sim_with_noise}
\end{figure}

In the second comparison case, colored measurement noises collected from the experiments are added. $\boldsymbol{R}_k$ is directly calculated using the applied noise values. Nevertheless, $\boldsymbol{Q}_k$ and $\boldsymbol{N}_k$ are difficult to tune because the uncertainties and gusts are far away from white noise. Their implemented values are $\boldsymbol{N}_k =10^{-5}\cdot [3.16,3.16,6.41,6.41,87.1,87.1]^\mathsf{T}$, $\boldsymbol{Q}_k = 1.02 \times 10^{-5}$. As shown in Table~\ref{table_GMLA_LQG_INDI} and Fig.~\ref{Dec_sim_with_noise}, mainly due to the phase lag induced by noise filtering, the load reduction rate of INDI-QP-V reduces to around 36~\% percent. The performance of LQG is even worse: the maximum value of $F_y - F_{y_*}$ is even amplified by 11.41~\% percent. As illustrated in Fig.~\ref{Dec_sim_INDI_LQR_GLMA_with_noise_servo}, although phase lag exists, INDI-QP-V actively makes the wing morph upwards to reduce the gust loads. By contrast, although LQG can alleviate the maneuver load by spanwise lift redistribution, it is not effective in alleviating the gust loads. To obtain better gust load alleviation performance, LQG has to be used along with some additional disturbance estimators (e.g., disturbance observer~\cite{Ferrier2018}).

\begin{figure}[!h]
\centering
\scalebox{1}{
\begin{subfigure}[t]{0.49\textwidth}
\includegraphics[width=\textwidth]{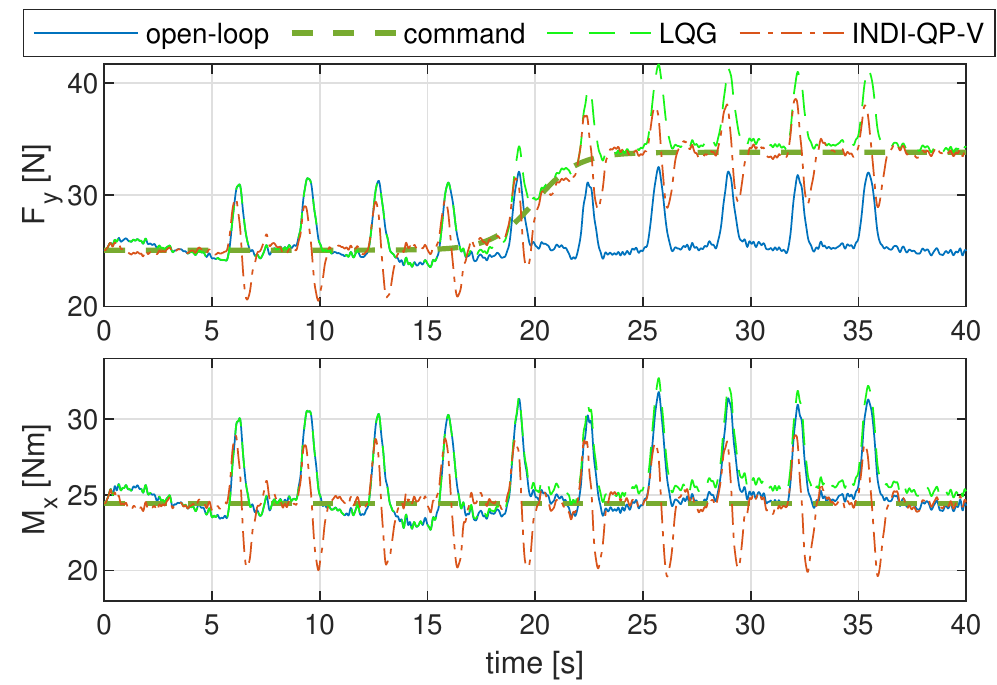}
\caption{Load responses.}
\label{Dec_sim_INDI_LQR_GLMA_with_everything_loads}
\end{subfigure}
\begin{subfigure}[t]{0.49\textwidth}
\includegraphics[width=\textwidth]{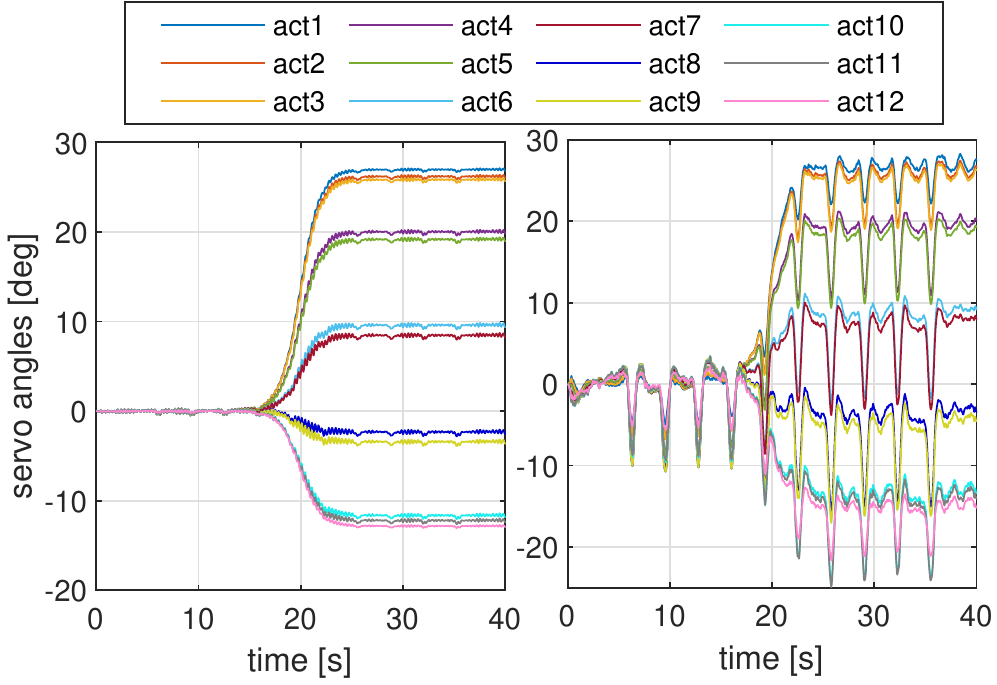}
\caption{Servo angles: LQG (left) and INDI-QP-V (right).}
\label{Dec_sim_INDI_LQR_GLMA_with_everything_servo}
  \end{subfigure}
}
\caption{Simultaneous gust and maneuver load alleviation with noise, fault, and backlash.}
\label{Dec_sim_everything}
\end{figure}

Apart from colored noises, actuator fault and backlash are also added to the last comparison case. Owing to the pick-up point failure, the 9$^{\text{th}}$ actuator effectiveness equals zero, while the 8$^{\text{th}}$ and 10$^{\text{th}}$ actuator effectiveness are respectively reduced by 53.20~\% and 26.52~\%. Equation~\eqref{backlash} is used to model backlash, with $k_1 = k_2 =1$, $u_{f_+} =-u_{f_-}= 0.6$~deg. Figure~\ref{Dec_sim_everything} and Table~\ref{table_GMLA_LQG_INDI} show that the performance of LQG is further degraded by the fault and backlash. Although the reduction rate of rms$(F_y - F_{y_*})$ is still positive under LQG control, the other performance metrics all become negative. On the contrary, INDI-QP-V can simultaneously alleviate gust and maneuver loads in spite of colored noises, actuator fault, and backlash. The rms value of $F_y - F_{y_*}$ is reduced by 68.08~\%, which is very close to the experimental result (67.76~\%). Under INDI-QP-V control, all the load metrics are reduced by over 19~\% in the simulation, and are alleviated by more than 44~\% in the experiment.

\section{Conclusions}
\label{sec_conclusions}
This paper presents the design and wind tunnel testing of a simultaneous gust and maneuver load alleviation control law for a seamless active morphing wing. To begin with, the incremental nonlinear dynamic inversion (INDI) control is derived for a generic multi-input/multi-output nonlinear system with an arbitrary relative degree. Then the closed-loop stability under the perturbation of model uncertainties, external disturbances, and control allocation errors are analyzed using Lyapunov methods. Moreover, two control allocation methods and their corresponding stability criteria are derived for INDI control. Although the INDI with pseudo inverse control allocation (INDI-PI) provides the least-squares solution, the input constraints are not considered. On the contrary, the actuator position constraints, rate constraints, and relative position constraints can all be satisfied by INDI with quadratic programming (INDI-QP). Furthermore, INDI-QP is augmented with the virtual shape functions (denoted as INDI-QP-V) to ensure the smoothness of the morphing wing. 

The effectiveness of the proposed INDI-QP-V has been validated by wind-tunnel experiments. During the experiment, the pick-up point of the 9$^{\text{th}}$ actuator was broken; the morphing mechanisms also presented unexpected hysteresis backlash behaviors. Despite these challenges, experimental results show that INDI-QP-V is robust to aerodynamic uncertainties, gusts, actuator faults, and nonlinear backlash. In maneuver load alleviation tasks, INDI-QP-V increased the total lift for performing pull-up maneuvers without amplifying the wing root bending moment. In the presence of successive ``1-cos'' gusts, INDI-QP-V mitigated the loads without requiring any gust information. Furthermore, INDI-QP-V made the seamless wing morph actively to modify the lift distribution and resist gusts at the same time. In all the tested cases, the input constraints were satisfied; the wing shape was smooth; the control law was realized in real time.

To further demonstrate the features of INDI-QP-V, is has been compared to the linear quadratic Gaussian (LQG) control. As a linear control method, LQG has to be used along with the tedious gain-scheduling method for nonlinear control problems. Its robust stability to model uncertainties and external disturbances is also not guaranteed. On the contrary, INDI-QP-V is a nonlinear control method with inherent robustness against uncertainties and disturbances. Moreover, INDI-QP-V has less model dependency, which simplifies its hardware implementation process. Furthermore, in simultaneous gust and maneuver load alleviation tasks, INDI-QP-V can more effectively alleviate the excessive loads. 

In conclusion, simulations and wind tunnel experiments have demonstrated that the proposed INDI-QP-V control is easy to implement, robust to actuator fault and backlash, and effective in simultaneously alleviating the gust and maneuver loads of the seamless active morphing wing.


\section*{Appendix}
\label{sec_appen}

\noindent\textit{\textbf{Proof for Theorem~\ref{theorem_1}}}: 

Denote the initial time point as $t_*$. Choose $V_1(\boldsymbol e) = \boldsymbol e^\mathsf{T} \boldsymbol  P \boldsymbol e$, where $\boldsymbol P = \boldsymbol P^\mathsf{T} >0$ is the solution of the Lyapunov equation $\boldsymbol P (\boldsymbol A_c - \boldsymbol B_c \boldsymbol K)+ (\boldsymbol A_c - \boldsymbol B_c \boldsymbol K)^\mathsf{T} \boldsymbol P = - \boldsymbol I$. Then $\alpha_1(\|\boldsymbol e\|_2) \leq  V_1 (\boldsymbol e) \leq   \alpha_2(\|\boldsymbol e\|_2),~
\alpha_1(\|\boldsymbol e\|_2) \triangleq \lambda_{\text{min}}(\boldsymbol P)\|\boldsymbol e\|_2^2,~\alpha_2(\|\boldsymbol e\|_2) \triangleq \lambda_{\text{max}}(\boldsymbol P)\|\boldsymbol e\|_2^2$. $\alpha_1,\alpha_2$ belong to the class $\mathcal K_{\infty} $ functions. Using Eq.~\eqref{closed_loop_indi}, the time derivative $V_1$ is:
\begin{eqnarray}
\dot V_1 &=& \boldsymbol e^\mathsf{T} [\boldsymbol P (\boldsymbol A_c - \boldsymbol B_c \boldsymbol K)+ (\boldsymbol A_c - \boldsymbol B_c \boldsymbol K)^\mathsf{T} \boldsymbol P] \boldsymbol e +  2\boldsymbol e^\mathsf{T} \boldsymbol P \boldsymbol B_c  \boldsymbol \varepsilon_{\text{indi}}  \nonumber \\
&\leq& -\|\boldsymbol e\|_2^2 + 2 \|\boldsymbol e\|_2 \|\boldsymbol P \boldsymbol B_c\|_2  \bar \varepsilon \leq - \theta_1  \|\boldsymbol e\|_2^2 , 
~~~ \forall \|\boldsymbol e\|_2 \geq \frac{2 \|\boldsymbol P \boldsymbol B_c\|_2  \bar \varepsilon}{1-\theta_1} \triangleq \mu_1  \bar \varepsilon
\label{V_bound}
\end{eqnarray}
with constant $\theta_1 \in (0,1)$. Consequently, for $\forall ~ \boldsymbol e (t_*) \in \mathbb{R}^{\rho}$, there exists a class $\mathcal {KL}$ function $\beta$ and finite $T_1 \geq 0$ such that $ \|\boldsymbol e(t)\|_2 \leq \beta(\|\boldsymbol e(t_*)\|_2,t-t_*), ~ t_* \leq \forall~ t\leq t_* +T_1,~ \|\boldsymbol e(t)\|_2 \leq \alpha_1^{-1}(\alpha_2(\mu_1 \bar \varepsilon)), ~ \forall ~t \geq t_*+T_1 \triangleq t_*'$. In other words, the tracking error $\boldsymbol e$ is bounded for all $t\geq t_*$ and is ultimately bounded by $\alpha_1^{-1}(\alpha_2(\mu_1\bar \varepsilon)) = \sqrt{\lambda_{\text{max}}(\boldsymbol P)/\lambda_{\text{min}}(\boldsymbol P)} \mu_1 \bar \varepsilon$. 

Regarding the internal dynamics, because the origin of $\dot {\boldsymbol \eta } =  \boldsymbol f_\eta (\boldsymbol \eta,\boldsymbol 0,\boldsymbol 0)$ is globally exponentially stable, then there exists a Lyapunov function $V_2(\boldsymbol \eta)$ defined in $D_{\eta} = \{\boldsymbol \eta \in \mathbb{R}^{n-\rho} \}$ that satisfies $ c_1 \|\boldsymbol \eta\|_2^2 \leq V_2(\boldsymbol \eta) \leq  c_2 \|\boldsymbol \eta\|_2^2$, $\frac{\partial V_2}{\partial \boldsymbol \eta} \boldsymbol f_\eta (\boldsymbol \eta, \boldsymbol 0,\boldsymbol 0) \leq  -c_3 \|\boldsymbol \eta\|_2^2$, $\quad \norm{\frac{\partial V_2}{\partial \boldsymbol \eta}}_2  \leq c_4 \|\boldsymbol \eta\|_2$, for some positive constants $c_1,c_2,c_3,c_4$. Denote $\alpha_1'(\|\boldsymbol \eta\|_2) \triangleq c_1  \|\boldsymbol \eta\|_2^2$, $\alpha_2'(\|\boldsymbol \eta\|_2) \triangleq  c_2\|\boldsymbol\eta\|_2^2$, then $\alpha_1', \alpha_2'$ belong to class $\mathcal{K}_{\infty}$ functions. Furthermore, because $ \boldsymbol f_\eta (\boldsymbol \eta,\boldsymbol \xi,\boldsymbol d)$ is continuously differentiable and globally Lipschitz in $(\boldsymbol \eta,\boldsymbol \xi,\boldsymbol d)$, then there exists a global Lipschitz constant $L$ such that $\|\boldsymbol f_\eta(\boldsymbol \eta,\boldsymbol \xi, \boldsymbol d) - \boldsymbol f_\eta (\boldsymbol \eta,\boldsymbol 0,\boldsymbol 0)\|_2 \leq L (\|\boldsymbol e\|_2 + \|\boldsymbol {\mathcal R} \|_2 + \|\boldsymbol d\|_2)$, $\forall \boldsymbol \eta  \in \mathbb{R}^{n-\rho} $. As a result, the time derivative of $V_2(\boldsymbol \eta)$ satisfies: 
\begin{eqnarray}
\dot V_2(\boldsymbol \eta) &=& \frac{\partial V_2}{\partial \boldsymbol \eta} \boldsymbol f_\eta(\boldsymbol \eta, \boldsymbol 0,\boldsymbol 0) + \frac{\partial V_2}{\partial \boldsymbol \eta}  [  \boldsymbol f_\eta (\boldsymbol \eta,\boldsymbol \xi, \boldsymbol d) - \boldsymbol f_\eta (\boldsymbol \eta,\boldsymbol 0,\boldsymbol 0) ]  \nonumber \\
&\leq&  -c_3 \|\boldsymbol \eta\|_2^2 +  c_4 L \|\boldsymbol \eta\|_2 (\|\boldsymbol e\|_2  + \bar{\mathcal R} +\bar{d}) 
\leq -c_3 (1-\theta_2) \|\boldsymbol \eta\|_2^2,~~~ \forall\|\boldsymbol \eta\|_2  \geq \frac{c_4L (\|\boldsymbol e\|_2  + \bar{\mathcal R} +\bar{d})}{c_3\theta_2} 
\label{eta_V2_eta}
\end{eqnarray}
with constant $\theta_2 \in (0,1)$. Denote
\begin{equation}
\mu_2 \triangleq \frac{c_4L (\sup_{t_*'\leq \tau \leq t}
\|\boldsymbol e\|_2  + \bar{\mathcal R} +\bar{d})}{c_3\theta_2} \triangleq \theta_3 (\sup_{t_*'\leq \tau \leq t}
\|\boldsymbol e\|_2  + \bar{\mathcal R} +\bar{d})
\label{mu_delta}
\end{equation}
then $\dot V_2(\boldsymbol \eta) \leq  -c_3 (1-\theta_2) \|\boldsymbol \eta\|_2^2,~ \forall\|\boldsymbol \eta\|_2 \geq \mu_2,~\forall t \geq t_*'$. Consequently, there exists a class $\mathcal{KL}$ function $\beta'$ such that $ \|\boldsymbol \eta(t)\|_2  \leq  \beta' (\|\boldsymbol \eta (t_*')\|_2,t-t_*') + \alpha_1'^{-1}(\alpha_2'(\mu_2)) ,~\forall t
\geq t_*'$. Since $ \beta'$ is a $\mathcal{KL}$ function, then the norm value of $\boldsymbol \eta(t)$ yields $ \|\boldsymbol \eta(t)\|_2  \leq  \theta_4 \bar{\varepsilon} +  \alpha_1'^{-1}(\alpha_2'(\theta_3 ( \alpha_1^{-1}(\alpha_2(\mu_1 \bar \varepsilon)) + \bar{\mathcal R} +\bar{d})  )),~\forall t
\geq t_*  +T_1+T_2$ for some finite $T_2>0$ and $ \theta_4>0$. In other words, $\boldsymbol \eta$ is globally ultimately bounded by a class $\mathcal{K}$ function of $\bar \varepsilon$, $\bar{\mathcal R}$, and $\bar d$.\hfill$\square$

\noindent\textit{\textbf{Proof for Theorem~\ref{theorem_2}}}: 

Essentially, Theorem~\ref{theorem_2} is a local version of Theorem~\ref{theorem_1}. When global Lipschitz and global exponential stability are not ensured, the stability criteria impose constraints on both initial condition and perturbation bound. Because the conditions for tracking error remain unchanged, Eq.~\eqref{V_bound} still holds, which proves that $\boldsymbol e$ is ultimately bounded by a class $\mathcal{K}$ function of $\bar \varepsilon$. Nevertheless, a $V_2(\boldsymbol \eta)$ and a Lipschitz constant only exist in a neighborhood of $\boldsymbol{\eta} = \boldsymbol{0}$, which is denoted as $D'_{\eta} = \{\boldsymbol \eta \in \mathbb{R}^{n-\rho} | ~\|\boldsymbol \eta\|_2 <r_{\eta} \}$. Take $0<r<r_{\eta} $ such that $D_r \subset D'_{\eta}$. According to the boundedness theories~\cite{Khalil}, Eq.~\eqref{eta_V2_eta} only holds when $\mu_2 < \alpha_2'^{-1}(\alpha_1'(r)),~\|\boldsymbol \eta (t_*')\|_2 \leq  \alpha_2'^{-1}(\alpha_1'(r)) $. Using Eq.~(\ref{mu_delta}), the perturbation is constrained by $\bar {\varepsilon} <  {\varepsilon}^* \triangleq (1/\mu_1) \alpha_2^{-1}( \alpha_1 ((1/\theta_3) ( \alpha_2'^{-1}(\alpha_1'(r))) -   \bar{\mathcal R} -\bar{d}))$. When the constraints on the initial condition and perturbation bound are satisfied, $\boldsymbol \eta$ is ultimately bounded by a class $\mathcal{K}$ function of $\bar \varepsilon$, $\bar{\mathcal R}$, and $\bar d$. \hfill$\square$

\noindent\textit{\textbf{Proof for Theorem~\ref{theorem_3}}}:
Recall Eqs.~(\ref{chp3_y_rho_expand},~\ref{u_indi_track},~\ref{closed_loop_indi}), the output dynamics under INDI control can also be written as $\boldsymbol y^{(\boldsymbol \rho)} = \boldsymbol  \nu_c +\boldsymbol \varepsilon_{\text{indi}}$. Also, at the previous time step, $\boldsymbol y^{(\boldsymbol \rho)}_0 = \boldsymbol  \nu_{c_0} +\boldsymbol \varepsilon_{\text{indi}_0}$. Therefore, using Eq.~(\ref{closed_loop_indi}), $\boldsymbol \varepsilon_{\text{indi}}$ can be rewritten as 
\begin{eqnarray}
\boldsymbol \varepsilon_{\text{indi}} &=& ({\boldsymbol{\mathcal{B}}}(\boldsymbol x_0) \bar{\boldsymbol{\mathcal{B}}}^{+}(\boldsymbol x_0)  - \boldsymbol I_{p\times p})(\boldsymbol  \nu_c -\boldsymbol y^{(\boldsymbol \rho)}_0 ) +  \boldsymbol \delta(\boldsymbol x, \Delta t)  + \boldsymbol \varepsilon_{\text{ca}} + \Delta \boldsymbol d_y \nonumber \\
&=&  (\boldsymbol I_{p\times p}-  {\boldsymbol{\mathcal{B}}}(\boldsymbol x_0) \bar{\boldsymbol{\mathcal{B}}}^{+}(\boldsymbol x_0)  )\boldsymbol \varepsilon_{\text{indi}_0} - (  \boldsymbol I_{p\times p}-{\boldsymbol{\mathcal{B}}}(\boldsymbol x_0) \bar{\boldsymbol{\mathcal{B}}}^{+}(\boldsymbol x_0)  )(\boldsymbol  \nu_c -\boldsymbol  \nu_{c_0})    + \boldsymbol \delta(\boldsymbol x, \Delta t)  + \boldsymbol \varepsilon_{\text{ca}} + \Delta \boldsymbol d_y  \nonumber \\
&\triangleq&  \boldsymbol E\boldsymbol \varepsilon_{\text{indi}_0} -\boldsymbol E \Delta \boldsymbol \nu_c + \boldsymbol \delta(\boldsymbol x, \Delta t)  + \boldsymbol \varepsilon_{\text{ca}} + \Delta \boldsymbol d_y 
\label{error_indi}
\end{eqnarray}
which can be written in a recursive way as $
\boldsymbol \varepsilon_{\text{indi}} (k) = \boldsymbol E(k) \boldsymbol \varepsilon_{\text{indi}} (k-1) - \boldsymbol E(k) \Delta \boldsymbol \nu_c (k) + \boldsymbol  \delta (k) +\boldsymbol \varepsilon_{\text{ca}} (k) +\Delta \boldsymbol d_y  (k) $. When the input constraints are not considered, the control allocation error $\boldsymbol \varepsilon_{\text{ca}}$ equals zero. Moreover, $\boldsymbol \nu_c$ is designed to be continuous in time (Eq.~(\ref{u_indi_track})), thus $\lim_{\Delta t \rightarrow 0} \|\boldsymbol \nu_c - \boldsymbol \nu_{c_0}\|_2 = 0,~\forall \boldsymbol x \in \mathbb{R}^n$. This equation also indicates that $\forall ~\overline{\Delta \nu}_c>0$, $\exists ~ \overline{\Delta t}>0, s.t.$ for all $ 0<\Delta t\leq \overline{\Delta t}$, $\forall \boldsymbol x\in \mathbb{R}^n,~\| \boldsymbol \nu_c - \boldsymbol \nu_{c_0}\|_2 \leq \overline{\Delta \nu}_c$. As a consequence, the following equation holds:
\begin{eqnarray}
\|\boldsymbol \varepsilon_{\text{indi}} (k) \|_2
&\leq& (\bar{b})^k \|\boldsymbol \varepsilon_{\text{indi}} (t=0) \|_2+ \sum_{j=1}^k (\bar{b})^{k-j+1} \| \Delta\boldsymbol \nu_c (j)\|_2 + \sum_{j=1}^{k-1} (\bar{b})^{k-j} \|\boldsymbol \delta (j) + \Delta \boldsymbol d (j)\|_2 + \|\boldsymbol \delta (k) + \Delta \boldsymbol d (k)\|_2 \nonumber \\
&\leq& (\bar{b})^k \|\boldsymbol \varepsilon_{\text{indi}} (t=0) \|_2+ \overline{\Delta \nu}_c \frac{\bar b - \bar b^{k+1}}{1- \bar b} + (\bar \delta + \overline{\Delta {d}}) \frac{1-\bar b^k}{1-\bar b}
\label{chp3_iterate_epsilon}
\end{eqnarray}

Since $\bar b <1$, Eq.~(\ref{chp3_iterate_epsilon}) satisfies $\|\boldsymbol \varepsilon_{\text{indi}} \|_2 \leq \frac{\overline{\Delta \nu}_c \bar{ b} + \bar \delta + \overline{\Delta {d}}}{1-\bar{b}}, ~\text{ as}~ k\rightarrow \infty$. In conclusion, $\boldsymbol \varepsilon_{\text{indi}}$ is bounded for all $k$, and is ultimately bounded by $\frac{\overline{\Delta \nu}_c \bar{b} + \bar \delta + \overline{\Delta {d}}}{1-\bar{b}}$.  
\hfill$\square$


\noindent\textit{\textbf{Proof for Theorem~\ref{theorem_4}}}: 

In contrast to Eq.~\eqref{u_indi_pi}, the analytical expression for the control increment given by quadratic programming $\Delta \boldsymbol{u}_{\text{indi-qp}}$ is unknown. Instead, the only information about $\Delta \boldsymbol{u}_{\text{indi-qp}}$ is that it satisfies $\bar{ {\boldsymbol{\mathcal{B}}}}(\boldsymbol x_0) \Delta \boldsymbol{u}_{\text{indi-qp}}  =  \boldsymbol \nu_c -\boldsymbol y^{(\boldsymbol \rho)}_0 + \boldsymbol \varepsilon_{\text{ca}}$, where $\boldsymbol \varepsilon_{\text{ca}}$ is the control allocation error. Using Eqs.~(\ref{chp3_y_rho_expand},~\ref{u_indi_track},~\ref{closed_loop_indi}), the corresponding $\boldsymbol \varepsilon_{\text{indi}}$ is derived as
\begin{equation}
\boldsymbol \varepsilon_{\text{indi}} = ( \boldsymbol{K}_{\mathcal{B}} (\boldsymbol{x}_0)- \boldsymbol{I}_{p\times p})( \boldsymbol \nu_c - \boldsymbol  \nu_{c_0} -\boldsymbol \varepsilon_{\text{indi}_0} + \boldsymbol \varepsilon_{\text{ca}}) + 
\boldsymbol \delta(\boldsymbol x, \Delta t) + 
\boldsymbol \varepsilon_{\text{ca}} + \Delta \boldsymbol d_y  
\triangleq \boldsymbol E'\boldsymbol \varepsilon_{\text{indi}_0} -\boldsymbol E' (\Delta \boldsymbol \nu_c+ \boldsymbol \varepsilon_{\text{ca}}) + \boldsymbol \delta(\boldsymbol x, \Delta t)  + \boldsymbol \varepsilon_{\text{ca}} + \Delta \boldsymbol d_y 
\end{equation}
which can be written in a recursive way as $
\boldsymbol \varepsilon_{\text{indi}} (k) = \boldsymbol E'(k) \boldsymbol \varepsilon_{\text{indi}} (k-1) - \boldsymbol E'(k) (\Delta \boldsymbol \nu_c (k) +\boldsymbol \varepsilon_{\text{ca}} (k)) + \boldsymbol  \delta (k) +\boldsymbol \varepsilon_{\text{ca}} (k) +\Delta \boldsymbol d_y  (k) $. Analogous to the proof of Theorem~\ref{theorem_3}, given a non-zero but bounded $\boldsymbol \varepsilon_{\text{ca}}$, the resulting $\boldsymbol \varepsilon_{\text{indi}}$ is bounded for all $k$, and is ultimately bounded by $\frac{\overline{\Delta \nu}_c \bar{b}' + \bar \delta + \overline{\Delta {d}} + (\bar{b}' + 1)\bar{\varepsilon}_{\text{ca}}}{1-\bar{b}'}$.  
\hfill$\square$

\noindent\textit{\textbf{Proof for Corollary~\ref{corollary_1}}}: 

Denote the solution of the INDI control with quadratic programming control allocation considering virtual shapes as $\Delta \boldsymbol{u}_{\text{indi-qp-v}}$, then correspondingly, $\boldsymbol \varepsilon_{\text{indi}}$ is $ \boldsymbol \varepsilon_{\text{indi}} = \boldsymbol \delta(\boldsymbol x, \Delta t)  + (\boldsymbol{\mathcal{B}}(\boldsymbol x_0)  \boldsymbol {\Phi}_{\bar{x}_s}- \bar{ {\boldsymbol{\mathcal{B}}}}(\boldsymbol x_0) \boldsymbol {\Phi}_{\bar{x}_s}) \Delta \boldsymbol u_{\text{indi-qp-v}} + 
\boldsymbol \varepsilon_{\text{ca}} + \Delta \boldsymbol d_y $. It is known that the control allocation leads to $\left(\bar{ {\boldsymbol{\mathcal{B}}}}(\boldsymbol x_0) \boldsymbol {\Phi}_{\bar{x}_s}\right)\Delta \boldsymbol{u}_{\text{indi-qp-v}}  =  \boldsymbol \nu_c -\boldsymbol y^{(\boldsymbol \rho)}_0 + \boldsymbol \varepsilon_{\text{ca}}$, thus:
\begin{equation}
\boldsymbol \varepsilon_{\text{indi}} 
= ( \boldsymbol{K}_{\mathcal{B}} (\boldsymbol{x}_0)- \boldsymbol{I}_{p\times p})( \boldsymbol \nu_c -\boldsymbol y^{(\boldsymbol \rho)}_0 + \boldsymbol \varepsilon_{\text{ca}}) + 
\boldsymbol \delta(\boldsymbol x, \Delta t) + 
\boldsymbol \varepsilon_{\text{ca}} + \Delta \boldsymbol d_y  
\end{equation}

The remaining derivations naturally follow the proof of Theorem~\ref{theorem_4}. Therefore, it is concluded that $\boldsymbol \varepsilon_{\text{indi}}$ is bounded for all $k$, and is ultimately bounded by $\frac{\overline{\Delta \nu}_c \bar{b}' + \bar \delta + \overline{\Delta {d}} + (\bar{b}' +1)\bar{\varepsilon}_{\text{ca}}}{1-\bar{b}'}$.  
\hfill$\square$


\begin{spacing}{1.0}
\bibliography{Ref_SmartX}
\end{spacing}
 
\end{document}